\documentclass[aps,prd,showpacs,onecolumn,preprintnumbers,notitlepage,superscriptaddress,
nofootinbib,amsfont,amssymb,10pt,singlespacing] {revtex4-1}
\usepackage{amsmath,bm}
\usepackage{times}
\usepackage{braket}
\usepackage{color,graphicx}
\usepackage{slashed}
\usepackage{hyperref}

\newcommand{\beq}{\begin{eqnarray}}
\newcommand{\eeq}{\end{eqnarray}}
\newcommand{\non}{\nonumber\\ }
\newcommand{\ov}{\overline}

\newcommand{\acp}{{\cal A}_{CP}}
\newcommand{\pb}{\phi_{B}}
\newcommand{\psl}{ P \hspace{-2.8truemm}/ }

\newcommand{\epsl}{\epsilon \hspace{-1.8truemm}/\,  }

\def\lsim{ {\ \lower-1.2pt\vbox{\hbox{\rlap{$<$}\lower6pt\vbox{\hbox{$\sim$}
}}}\ } }
\def\gsim{ {\ \lower-1.2pt\vbox{\hbox{\rlap{$>$}\lower6pt\vbox{\hbox{$\sim$}
}}}\ } }

\def \jhep{ J. High Energy Phys.  }

\definecolor{Red}{rgb}{1.,0.,0.}

\definecolor{Blue}{rgb}{0.,0.,1.}

\definecolor{nicered}{rgb}{0.7,0.1,0.1}
\definecolor{nicegreen}{rgb}{0.1,0.5,0.1}

\hypersetup{colorlinks,citecolor=nicegreen,linkcolor=nicered}

\begin{document}

\title{ %Branching ratios, polarization fractions, and CP-violating asymmetries\\
Penguin-dominated $B \to \phi K_1(1270)$ and $ %B \to
\phi K_1(1400) $ decays in the perturbative QCD approach}
%%%==================================================================

\author{Xin~Liu}
\email[Electronic address:]{liuxin.physics@gmail.com}
\affiliation{School of Physics and Electronic Engineering,\\ Jiangsu Normal University, Xuzhou, Jiangsu 221116,
People's Republic of China}

\author{Zhi-Tian~Zou}
\email[Electronic address:]{zouzt@ytu.edu.cn}
\affiliation{Department of Physics, Yantai University, Yantai, Shandong 264005, People's Republic of China}

\author{Zhen-Jun~Xiao}
\email[Electronic address:]{xiaozhenjun@njnu.edu.cn}
\affiliation{Department of Physics and Institute of Theoretical
Physics,\\ Nanjing Normal University, Nanjing, Jiangsu 210023,
People's Republic of China}

\date{\today}

%%%%%%%%%%%%%%%%%%%%%%%%%%%%%%%%%%%%%%%%%%%%%%%%%%%%%%%%%%%%%%%%%%
\begin{abstract}

We investigate the CP-averaged branching ratios, the polarization fractions,
the relative phases, and the CP-violating asymmetries of the penguin-dominated
$B \to \phi K_1(1270) $ and $ \phi  K_1(1400)$ decays in
the perturbative QCD(pQCD) approach,
where $K_1(1270)$ and $K_1(1400)$ are believed to be the
mixtures of two distinct types of axial-vector $K_{1A}(^3P_1)$ and
$K_{1B}(^1P_1)$ states with different behavior, however, their
mixing angle $\theta_{K_1}$ is still a hot and controversial topic presently.
By numerical evaluations with two different mixing angles $\theta_{K_1} \sim 33^\circ$
and $58^\circ$ and phenomenological analysis,
we find that:
(a) the pQCD predictions for the branching ratio, the longitudinal polarization fraction
and the direct CP violation of $B^\pm \to \phi K_1(1270)^\pm$ decay with the smaller angle
$33^\circ$ are in good agreement with the currently available data;
(b) though the central values significantly exceed the available upper limit,
both pQCD predictions of $Br(B^\pm \to \phi K_1(1400)^\pm)$ with two different mixing angles
are consistent with that obtained in QCD factorization and with the preliminary data in 2$\sigma$ errors.
These results and other relevant predictions for the considered decays
will be further tested by the LHCb and the forthcoming Super-B experiments;
(c) the weak annihilation contributions can play an important role in $B \to \phi K_1(1270)$
and $\phi K_1(1400)$ decays;
(d) these pQCD predictions combined with the future precision measurements can
examine the reliability of the factorization approach employed here, but also explore
the complicated QCD dynamics and mixing angle $\theta_{K_1}$ of
the axial-vector $K_1(1270)$ and $K_1(1400)$ system.

\end{abstract}
%%%%%%%%%%%%%%%%%%%%%%%%%%%%%%%%%%%%%%%%%%%%%%%%%%%%%%%%%%%%%%

\pacs{13.25.Hw, 12.38.Bx, 14.40.Nd}
%\preprint{\footnotesize JSNU-PHY-TH-2014}
\maketitle

%
%%%
%%%%%%%%%%%%%%%%% I. INTRODUCTION %%%%%%%%%%%%%%%%%%%%%%%%%%%%%%%%
%%%
%

\section{Introduction}

In the quark model, the possible quantum numbers $J^{PC}$ for the
orbitally excited axial-vector mesons are $1^{++}$ or $1^{+-}$,
depending on different spin couplings of the involved two quarks. In the
SU(3) limit, those mesons can not mix with each other; but, since the
$s$ quark is heavier than $u,d$ quarks, the physical states of strange axial-vector mesons, $K_1(1270)$ and $K_1(1400)$, are believed to be
mixtures of two distinct types of $K_{1A}$
and $K_{1B}$, where $K_{1A}$ and $K_{1B}$ are
$^3P_1$ and $^1P_1$ states, respectively.
Because both $K_1$(Hereafter, for the sake of simplicity,
we will adopt $K_1$ to denote $K_1(1270)$ and $K_1(1400)$
unless otherwise stated.) mesons are not pure $^3P_1$
or $^1P_1$ states, the mixing angle $\theta_{K_1}$
between two axial-vector states $K_{1A}$ and $K_{1B}$ is now of great interest at both theoretical and experimental aspects.
Furthermore, the mixing angle $\theta_{K_1}$ can be utilized to determine the
mixing angle $\theta_{^1P_1}$ and $\theta_{^3P_1}$ with the former(latter)
being the mixing angle of $h_1(1170)(f_1(1285))$ and $h_1(1380)(f_1(1420))$
in the flavor basis through mass relations(for detail, see recent
discussion~\cite{Cheng:2013cwa}).
However, this mixing angle $\theta_{K_1}$ is still an issue
in controversy presently.
It is therefore definitely interesting to investigate the mixing
angle $\theta_{K_1}$ through kinds of ways, for example, examining the hints of $\theta_{K_1}$
in the rare $B$ meson decays to the final states involving the aforementioned $K_1$ mesons.

Recently, the BABAR Collaboration has
measured the branching ratio, the longitudinal
polarization fraction and the
direct CP asymmetry(Here, the
definition of the direct CP asymmetry $\acp$ is $\frac{\Gamma^+
-\Gamma^-}{\Gamma^+ + \Gamma^-}$~\cite{Aubert:2008bc}, where
$\Gamma^+$ and $\Gamma^-$ denote the decay width of $B^+$ and $B^-$
meson, respectively.) of $B^{\pm} \to \phi K_1(1270)^{\pm}$
decay~\cite{Aubert:2008bc} for the first time,
\beq
Br(B^{\pm} \to \phi K_1(1270)^{\pm}) &=&
(6.1 \pm 1.6 \pm 1.1) \times 10^{-6}\;; \non
f_L(B^{\pm} \to \phi K_1(1270)^{\pm})&=&
0.46^{+0.12+0.06}_{-0.13-0.07}\;;\label{eq:exp1}\\
A_{CP}(B^{\pm} \to \phi K_1(1270)^{\pm})&=&
+0.15 \pm 0.19 \pm 0.05\;; \nonumber
\eeq
and placed the upper limit at 90\% C.L. on the branching
ratio of $B^{\pm} \to \phi K_1(1400)^{\pm}$
decay~\cite{Aubert:2008bc},
\beq
Br(B^{\pm} \to \phi K_1(1400)^{\pm})
&<& 3.2(0.3 \pm 1.6 \pm 0.7) \times 10^{-6}\;.\label{eq:exp2}
\eeq
One can easily observe that the vector-axial-vector $B^\pm \to  \phi K_1(1270)^\pm$ decay looks more like
the vector-vector $B^\pm \to \phi K^{*\pm}$ one, which has been confirmed experimentally that the transverse amplitudes
account for a large fraction~\cite{Aubert:2003mm,Aubert:2004xc,Aubert:2006uk,Aubert:2007ac,Chen:2003jfa,Chen:2005zv}.
The precision of above measurements for $B^\pm \to \phi K_1(1270)^\pm$
will be improved
rapidly in the relevant Large Hadron Collider beauty (LHCb) experiments.
Moreover, the observation
on the $B^0 \to \phi K_1(1270)^0$ and $B \to \phi K_1(1400)$ decays could also be made with
good precision at LHCb in the near future.

It is well known that the study of exclusive non-leptonic
weak decays of $B$ mesons provides
not only good opportunities for testing the standard model(SM) but
also powerful means for probing different new physics scenarios
beyond the SM. Just like the two body charmless hadronic $B \to
VV$ decays, $B \to VA $ modes are also expected to
have rich physics as they have three polarization states. Through polarization studies,
these channels can shed light on the
underlying helicity structure of the decay mechanism~\cite{Cheng:2008gxa}.
Experimentally,
measurements of polarization in rare vector-vector B
meson decay, such as $B \to \phi K^*$, have revealed an
unexpectedly large fraction of transverse polarization,
which violates the naively expected hierarchy, i.e., $f_L \sim 1$ and $f_\parallel \approx  f_\perp \sim {\cal O}(m_V^2/m_B^2)$, with
$f_L$, $f_\parallel$, and $f_\perp$ denoting the polarization fractions on longitudinal, parallel, and
perpendicular polarization, respectively.
In view of the polarization anomalies exhibited in $B \to \phi K^*$ decays
and the same transition pattern $\bar b \to \bar s s \bar s$ involved
in both $B \to \phi K^*$ and $B \to \phi K_1$ decays, it is of particular interest
to see whether similar anomalies occur in $B \to \phi K_1$ decays.
Moreover, in order to find out the real causes of the above mentioned
polarization anomalies in these types of decays,
it demands considerable studies on more processes.

At the theoretical aspect, up to now, the two-body hadronic $B \to \phi K_1$ decays
have been investigated by G.~Calder${\rm \acute{o}}$n {\it et al.}~\cite{Calderon:2007nw}
in naive factorization approach,
by Chen {\it et al.}~\cite{Chen:2005cx} in
generalized factorization approach(GFA), and by Cheng and Yang~\cite{Cheng:2008gxa}
in QCD factorization(QCDF), respectively. The theoretical predictions
were given with the mixing angle $\theta_{K_1} \approx 32^\circ, 58^\circ$
in Ref.~\cite{Calderon:2007nw} and $\theta_{K_1} \approx 37^\circ, 58^\circ$
in Refs.~\cite{Chen:2005cx,Cheng:2008gxa}.
However, those predictions of the decay rates and polarization
fractions for the considered $B \to \phi K_1$
decays presented very different phenomenologies.
For the case of  $B^+ \to \phi K_1(1270)^+$ decay, for instance,
the authors  of Ref.~\cite{Calderon:2007nw}
found that the branching ratio of the considered decay is in the order of $10^{-9} \sim 10^{-7}$,
which is much smaller than currently available data.
Furthermore, the numerical results were also very
sensitive to the variation of the mixing angle $\theta_{K_1}$:
$Br(B^+ \to \phi K_1(1270)^+) \sim 4 \times 10^{-9}$ or $ 3 \times 10^{-7}$ for
$\theta_{K_1} \approx 32^\circ$ or $58^\circ$, respectively.
The relevant polarization fractions were not evaluated in Ref.~\cite{Calderon:2007nw}.
By neglecting the so-called negligible annihilation contributions,
the authors  of Ref.~\cite{Chen:2005cx} predicted the branching ratio $Br(B^+ \to \phi K_1(1270)^+)\sim
10^{-5}$ with the preferred $N_c^{\rm eff} = 2$ or $3$, where $N_c^{\rm eff}$
was the effective color number containing the non-factorizable effects.
When $N_c^{\rm eff}$ is close to 5, the numerical results for the decay rate with
mixing angle $37^\circ$ and $58^\circ$ are well consistent with the present measurement.
Moreover, the calculations showed the moderate dependence on the mixing angle $\theta_{K_1}$ and preferred the smaller angle for the $B^+ \to \phi K_1(1270)^+$ mode. And the longitudinal
polarization fractions were predicted around 90\% in the cases
of both $37^\circ$ and $58^\circ$.
By using the light-cone QCD sum rule results for the
$B \to K_{1A}$ and $B \to K_{1B}$ form factors \cite{Cheng:2008gxa}, the authors predicted the
branching ratios and  polarization fractions in QCDF by adopting
the penguin-annihilation parameters inferred from the $B \to \phi K^*$ decays. The decay rate
for $B^+ \to \phi K_1(1270)^+$ mode is $(3.8^{+5.4}_{-3.4}) \times 10^{-6}$
with $\theta_{K_1} \approx 37^\circ$
and $(3.4^{+5.9}_{-3.2}) \times 10^{-6}$ with $\theta_{K_1} \approx 58^\circ$, respectively, which is consistent
with the data within errors and shows the weak dependence on the mixing angle $\theta_{K_1}$.
While the predicted longitudinal polarization
fractions exhibit the dominant longitudinal(transverse) contributions for $B^+ \to \phi K_1(1270)^+$
with mixing angle $\theta_{K_1} \approx 37^\circ(58^\circ)$.
Frankly speaking, the large discrepancies among those
predicted branching ratios and polarization fractions of the considered decays
indicate that more studies by employing new approaches and/or methods are
greatly needed to explore these decay modes and understand in depth the
physics hidden in them.

In this work, we will calculate the CP-averaged branching ratios,
the polarization fractions, the relative phases,
and the CP-violating asymmetries of the four charmless hadronic
$B \to \phi K_1$ decays~\footnote{These considered $B \to \phi K_1$ decays
are analogous to
vector-vector $B \to \phi K^*$ decays, which are expected to arise only from virtual
loop effects in the standard model and are particularly sensitive to the contributions
from beyond the standard model~\cite{Valencia:1988it,Datta:2003mj}.
Moreover, some physical quantities
such as the relative phases and the CP-violating asymmetries of $B \to \phi K_1$
decays are evaluated for the first time in this work.} by
employing the low energy effective Hamiltonian~\cite{Buchalla:1995vs}
and the perturbative QCD (pQCD)
factorization approach\cite{Keum:2000ph,Keum:2000wi,Lu:2000em,Li:2003yj} based on the $k_T$
factorization theorem.
By keeping the transverse momentum $k_T$ of the quarks, the pQCD approach is free of endpoint singularity
and the Sudakov formalism makes it more self-consistent.
In the pQCD approach, we can explicitly evaluate not only the factorizable and non-factorizable
spectator diagrams, but also the weak annihilation ones.
Although there is a different viewpoint on the evaluations of annihilation diagrams proposed
in the soft-collinear effective theory (See Refs.~\cite{Arnesen:2006vb,Chay:2007ep} for details),
the previous predictions on the annihilation contributions in heavy flavor
$B$ meson decays calculated with the pQCD approach
have already been tested at various aspects, for example, branching ratios of pure
annihilation $B_d \to D_s^- K^+$, $B_d \to K^+ K^-$, and $B_s \to \pi^+ \pi^-$
decays~\cite{Lu:2002iv,Li:2004ep,Ali:2007ff,Xiao:2011tx},
direct CP asymmetries of $B^0 \to \pi^+\pi^-$, $K^+\pi^-$
decays~\cite{Keum:2000ph,Keum:2000wi,Lu:2000em,Hong:2005wj}, and the
explanation of $B\to \phi K^*$ polarization
problem~\cite{Li:2004ti,Li:2004mp}, which indicate that the
pQCD approach is a reliable method to deal with the annihilation diagrams.

The paper is organized as follows. In Sec.~\ref{sec:form}, we present
the formalism, hadron wave functions and perturbative calculations
of the considered four $B \to \phi K_1$ decays.
The numerical results and the corresponding phenomenological
analyses are addressed in Sec.~\ref{sec:randd}. Finally,
Sec.~\ref{sec:summary} contains the main conclusions and a short summary.

%%%
%%%%%%%%%%%%%%%%% II. Formalism %%%%%%%%%%%%%%%%%%%%%%%%%%%%%%%%
%%%

\section{ Formalism}\label{sec:form}

The pQCD approach is one of the popular methods to evaluate
the hadronic matrix elements in the heavy $b$-flavor
mesons' decays. The basic idea of the pQCD approach is that it takes into
account the transverse momentum $k_T$ of the valence quarks
in the calculation of the hadronic matrix elements.
The $B$ meson transition form factors, and the non-factorziable spectator and
annihilation contributions are then all calculable in the framework
of the $k_T$ factorization, where three energy scales $m_W, m_B$ and
$t\approx \sqrt{m_B \Lambda_{\rm QCD}} $ are
involved~\cite{Keum:2000ph,Keum:2000wi,Lu:2000em,Li:1994cka,Li:1995jr,Li:1994iu}.
The running of the Wilson coefficients $C_i(t)$ with $t \geq \sqrt{m_B \Lambda_{\rm QCD}} $
are controlled by
the renormalization group equation and can be calculated perturbatively.
The dynamics below $\sqrt{m_B \Lambda_{\rm QCD}}$ is soft, which is described by the meson wave
functions. The soft dynamics is not perturbative but universal for all channels.

In the pQCD approach, the amplitude of $B \to \phi K_1$ decays can therefore be factorized
into the convolution of the six-quark hard kernel($H$), the jet function($J$) and the Sudakov
factor($S$) with the bound-state wave functions($\Phi$) as follows,
\begin{eqnarray}
A(B \to \phi K_1)=\Phi_{B} \otimes H \otimes J \otimes S
\otimes \Phi_{\phi} \otimes \Phi_{K_1}\;, \label{eq:sixquarks}
\end{eqnarray}
The function $\Phi$ is
the wave function describing hadronization of the quark and anti-quark to the meson,
which is independent of the specific processes and usually determined by employing
nonperturbative QCD techniques or other well measured processes.
The jet function $J$ comes from the threshold resummation, which
exhibits strong suppression effect in the small $x$ (quark momentum fraction)
region~\cite{Li:2001ay,Li:2002mi}.
The Sudakov factor $S$ comes from the $k_T$ resummation, which
provide a strong suppression in the small $k_T$ region~\cite{Botts:1989kf,Li:1992nu}.
These resummation effects therefore guarantee the removal of the endpoint singularities.

Because of the rather heavy $b$ quark, for convenience, we usually work in the rest frame of
$B$ meson. By utilizing the light-cone coordinate $(P^+, P^-, {\bf P}_T)$
to describe the meson's momenta with the definitions
\beq
P^{\pm} &=& \frac{p_0 \pm p_3}{\sqrt{2}} \qquad {\rm and} \qquad {\bf P}_T= (p_1, p_2)\;;
 \eeq
we can write the involved three meson momenta in the $B \to \phi K_1$ decays,
%Then for $B \to \phi K_1$ decays, the involved three meson momenta in the
%light-cone coordinates can be written as
\beq
     P_1=\frac{m_{B}}{\sqrt{2}} (1,1,{\bf 0}_T), \qquad
     P_2 =\frac{m_{B}}{\sqrt{2}} (1-r_3^2,r_2^2,{\bf 0}_T), \qquad
     P_3 =\frac{m_{B}}{\sqrt{2}} (r_3^2,1-r_2^2,{\bf 0}_T),
\eeq
respectively, where the $\phi$ ($K_1$) meson moves in the plus (minus) $z$ direction
carrying the momentum $P_2$ ($P_3$) and $r_2=m_{\phi}/m_B$, $r_3=m_{K_1}/m_{B}$.
When we choose the (light) quark momenta in $B$, $\phi$ and $K_1$ mesons as $k_1$,
$k_2$, and $k_3$, respectively, and define
\beq
k_1 = (x_1P_1^+,0,{\bf k}_{1T}), \quad k_2 = x_2 P_2+(0,0,{\bf k}_{2T}),
\quad k_3 = x_3 P_3+(0, 0,{\bf k}_{3T}).
\eeq
then integrate out $k_1^-$, $k_2^-$, and
$k_3^+$ in the Eq.(\ref{eq:sixquarks}), the more explicit form of the decay amplitude
for $B \to \phi K_1$ decays can be conceptually rewritten as the following,
\beq
A(B \to \phi K_1) &\sim &\int\!\! d x_1 d
x_2 d x_3 b_1 d b_1 b_2 d b_2 b_3 d b_3
\non && \cdot {\mathrm{Tr}}
\left [ C(t) \Phi_{B}(x_1, b_1) \Phi_{\phi}(x_2, b_2)
\Phi_{K_1}(x_3, b_3) H(x_i, b_i, t) S_t(x_i)\, e^{-S(t)} \right ].
\label{eq:a2}
\eeq
where $b_i$ is the conjugate space coordinate
of $k_{iT}$, and $t$ is the largest energy scale in hard kernel
$H(x_i,b_i,t)$. Tr denotes the trace over Dirac and color indices.
$C(t)$ stands for the Wilson coefficients including
the large logarithms $\ln (m_W/t)$. $S_t(x_i)$ and $e^{-S(t)}$ correspond
to the jet function $J$ and Sudakov factor $S$ in Eq.~(\ref{eq:sixquarks}),
respectively, whose detailed expressions can be easily found in the original Refs.~\cite{Li:2001ay,Li:2002mi,Botts:1989kf,Li:1992nu}. Thus, with Eq.~(\ref{eq:a2}),
we can give the convoluted amplitudes of the $B \to \phi K_1$ decays,
which will be presented in the next section, through
the evaluations of the hard kernel $H(x_i,b_i,t)$ at leading order in $\alpha_s$ expansion
in the pQCD approach.

%The longitudinal and transverse
%polarization vectors of axial-vector meson are denoted by $\epsilon^L$ and $\epsilon^T$, respectively, satisfying $P
%\cdot \epsilon=0$ in each polarization.
%The longitudinal polarization
%vectors, $\epsilon_2^L$ and $\epsilon_3^L$, can be chosen as
%\beq
%\epsilon_2^L &=& \frac{m_{B}}{\sqrt{2} m_{\phi}} (1-r_3^2, -r_2^2,{\bf
%0}_T) \qquad  {\rm and} \qquad \epsilon_3^L = \frac{m_{B}}{\sqrt{2} m_{K_1}} (-r_3^2,
%1-r_2^2,{\bf 0}_T).
%\eeq
%And the transverse ones are parameterized
%as
%$\epsilon_2^T = (0, 0, {\bf 1}_T)$
%and
%$\epsilon_3^T = (0, 0, {\bf 1}_T)$.

%Putting the (light) quark momenta in $B$, $\phi$ and $K_1$ mesons as $k_1$,
%$k_2$, and $k_3$, respectively, we define
%\beq
%k_1 = (x_1P_1^+,0,{\bf k}_{1T}), \quad k_2 = x_2 P_2+(0,0,{\bf k}_{2T}),
%\quad k_3 = x_3 P_3+(0, 0,{\bf k}_{3T}).
%\eeq
%Then, for $B \to \phi K_1$ decays, the integration over $k_1^-$, $k_2^-$, and
%$k_3^+$ will conceptually lead to the decay amplitude in the pQCD
%approach,

\subsection{Wave functions and distribution amplitudes}\label{ssec:wf}

%Throughout this paper, we will use light-cone coordinate
%$(P^+, P^-, {\bf P_T})$ to describe the meson's momenta
%with the definitions
%$P^{\pm}=(p_0 \pm p_3)/\sqrt{2}$ and ${\bf P_T}=(p_1,p_2)$.
The heavy $B$ meson is usually treated as a heavy-light system and its light-cone wave function
can generally be defined as~\cite{Keum:2000ph,Keum:2000wi,Lu:2000em,Lu:2002ny}
\beq
\Phi_{B,\alpha\beta,ij}&\equiv&
  \langle 0|\bar{b}_{\beta j}(0)q_{\alpha i}(z)|B(P)\rangle \non
&=& \frac{i \delta_{ij}}{\sqrt{2N_c}}\int dx d^2 k_T e^{-i (xP^-z^+ - k_T z_T)}
\left\{(\psl +m_{B})\gamma_5
 \phi_{B}(x, k_T) \right\}_{\alpha\beta}\;;
\label{eq:def-bq}
\eeq
where the indices $i,j$ and $\alpha,\beta$ are the Lorentz indices and color indices respectively,
$P(m)$ is the momentum(mass) of the $B$ meson, $N_c$ is the color factor, and
$k_T$ is the intrinsic transverse momentum of the light quark in $B$ meson.
%Note that, in principle, there are two Lorentz
%structures of the wave function to be considered in the numerical calculations,
%however,
%the contribution induced by the second Lorentz structure
%is numerically small and approximately negligible~\cite{Lu:2002ny}.

In Eq.~(\ref{eq:def-bq}), $\phi_{B}(x,k_T)$ is the $B$ meson distribution amplitude
and obeys to the following normalization condition,
\beq
\int_0^1 dx \phi_{B}(x, b=0) &=& \frac{f_{B}}{2 \sqrt{2N_c}}\;.\label{eq:norm}
\eeq
where $b$ is the conjugate space coordinate of transverse momentum $k_T$ and $f_B$
is the decay constant of $B$ meson.
For $B$ meson, the distribution amplitude in the impact $b$
space has been proposed
\beq
\phi_{B}(x,b)&=& N_Bx^2(1-x)^2
\exp\left[-\frac{1}{2}\left(\frac{xm_B}{\omega_b}\right)^2
-\frac{\omega_b^2 b^2}{2}\right] \;,
\eeq
in Refs.~\cite{Keum:2000ph,Keum:2000wi,Lu:2000em},
where the normalization factor $N_{B}$
is related to the decay constant $f_{B}$ through Eq.~(\ref{eq:norm}).
The shape parameter $\omega_b$ has been fixed at $\omega_b=0.40\pm 0.04$~GeV by using the rich experimental
data on the $B$ mesons with $f_{B}= 0.19$~GeV based on lots of calculations of form factors~\cite{Lu:2002ny}
and other well-known decay modes of $B$ mesons~\cite{Keum:2000ph,Keum:2000wi,Lu:2000em}
in the pQCD approach in recent years.

The light-cone wave functions of the vector meson $\phi$
and axial-vector state $K_{1A(B)}$
have been given
in the QCD sum rule method up to twist-3 as~\cite{Ball:1998sk,Ball:2007rt}
 \beq
\Phi^{L}_{\phi,\alpha\beta,ij}&\equiv&
\langle \phi(P, \epsilon_L)|\bar q(z)_{\beta j} q(0)_{\alpha i} |0\rangle \non
 &=&  \frac{ \delta_{ij}}{\sqrt{2 N_c}} \int^1_0dxe^{ix P\cdot
 z}  \biggl\{ m_{\phi}\, {\epsl}_L \,\phi_{\phi}(x)  +
 {\epsl}_L \, \psl\,\phi_{\phi}^t(x)  + m_{\phi}\, \phi_{\phi}^s(x) \biggr\}_{\alpha\beta}\;,
 \eeq
 \beq
\Phi^{T}_{\phi,\alpha\beta,ij}&\equiv&
\langle \phi(P, \epsilon_T)|\bar q(z)_{\beta j} q(0)_{\alpha i} |0\rangle \non
 &=&  \frac{ \delta_{ij}}{\sqrt{2 N_c}} \int^1_0dxe^{ix P\cdot
 z}  \biggl\{ m_{\phi}\, {\epsl}_T\, \phi_{\phi}^v(x) +
{\epsl}_T\, {\psl}_{\phi} \phi_{\phi}^T(x)+m_{\phi}
i\epsilon_{\mu\nu\rho\sigma}\gamma_5\gamma^\mu {\epsl}_T^{\nu}
n^\rho v^\sigma \phi_{\phi}^a(x) \biggr\}_{\alpha\beta}\;,
 \eeq
 and~\cite{Yang:2007zt,Li:2009tx}
 \beq
\Phi^{L}_{K_{1A(B)},\alpha\beta,ij}&\equiv&
\langle K_{1A(B)}(P, \epsilon_L)|\bar q(z)_{\beta j} q(0)_{\alpha i} |0\rangle \non
 &=&  \frac{ \delta_{ij} \gamma_5}{\sqrt{2 N_c}} \int^1_0dxe^{ix P\cdot
 z}  \biggl\{ m_{K_{1A(B)}}\, {\epsl}_L \,\phi_{K_{1A(B)}}(x)  +
 {\epsl}_L \, \psl\,\phi_{K_{1A(B)}}^t(x)  + m_{K_{1A(B)}}\, \phi_{K_{1A(B)}}^s(x) \biggr\}_{\alpha\beta}\;,
 \eeq
 \beq
\Phi^{T}_{K_{1A(B)},\alpha\beta,ij}&\equiv&
\langle K_{1A(B)}(P, \epsilon_T)|\bar q(z)_{\beta j} q(0)_{\alpha i} |0\rangle \non
 &=&  \frac{ \delta_{ij} \gamma_5}{\sqrt{2 N_c}} \int^1_0dxe^{ix P\cdot
 z}  \biggl\{ m_{K_{1A(B)}}\, {\epsl}_T\, \phi_{K_{1A(B)}}^v(x) +
{\epsl}_T\, {\psl}_{K_{1A(B)}} \phi_{K_{1A(B)}}^T(x)
 \non && \hspace{6.68cm}+m_{K_{1A(B)}}
i\epsilon_{\mu\nu\rho\sigma}\gamma_5\gamma^\mu {\epsl}_T^{\nu}
n^\rho v^\sigma \phi_{K_{1A(B)}}^a(x) \biggr\}_{\alpha\beta}\;,
\eeq
for longitudinal polarization and transverse polarization, respectively,
with the polarization vectors $\epsilon_L$ and $\epsilon_T$ of $\phi$
or $K_{1A(B)}$,
satisfying $P \cdot \epsilon=0$, where
$x$ denotes the momentum
fraction carried by quark in the meson, and $n=(1,0,{\bf 0}_T)$
and $v=(0,1,{\bf 0}_T)$ are dimensionless light-like unit vectors.
We adopt the convention $\epsilon^{0123}=1$ for the
Levi-Civita tensor $\epsilon^{\mu\nu\alpha\beta}$.

The twist-2 distribution amplitudes $\phi_{\phi}$ and $\phi_{\phi}^T$
can be parameterized as:
\beq
\phi_{\phi}(x)&=&\frac{f_{\phi}}{2\sqrt{2N_c}}6x (1-x)
 \left[1+a_{2\phi}^{||} \frac{3}{2}(5 (2x -1)^2-1) \right],\\
\phi_{\phi}^T(x)&=&\frac{f_{\phi}^T}{2\sqrt{2N_c}}6x (1-x)
 \left[1+a_{2\phi}^{\perp} \frac{3}{2}(5 (2x -1)^2-1) \right],\label{phiV}
\eeq
Here $f_{\phi}$ and $f_{\phi}^T$ are the decay constants of the $\phi$ meson
with longitudinal and transverse polarization, respectively, whose values are~\cite{Beringer:1900zz,Ball:2004rg}
\beq
 f_{\phi} &=& 0.231 \pm 0.004 ~~~~~{\rm GeV}\;,  \qquad f_{\phi}^T = 0.200 \pm 0.010 ~~~~~~{\rm GeV}\;.
\eeq
The Gegenbauer moments $a_{2\phi}^{||,\perp}$
are mainly determined by the technique of QCD sum rules. Here we
quote the recent updates~\cite{Ball:2007rt,Ball:2004rg} as
\beq
 a_{2\phi}^\parallel & = & 0.18 \pm 0.08,\;\;\;\;\;    a_{2\phi}^\perp  =0.14 \pm
 0.07\;,
\eeq
where the values are taken at $\mu=1$ GeV.

The asymptotic forms of the twist-3 distribution amplitudes
$\phi^{t,s}_{\phi}$ and $\phi_\phi^{v,a}$ are adopted:
\beq
\phi^t_\phi(x) &=& \frac{3f^T_\phi}{2\sqrt {2N_c}}(2x-1)^2,\;\;\;\;\;\;\;\;\;\;\;
  \hspace{0.5cm} \phi^s_\phi(x)=-\frac{3f_\phi^T}{2\sqrt {2N_c}} (2x-1)~,\\
\phi_\phi^v(x)&=&\frac{3f_\phi}{8\sqrt{2N_c}}(1+(2x-1)^2),\;\;\;\;\; \ \
 \phi_\phi^a(x)=-\frac{3f_\phi}{4\sqrt{2N_c}}(2x-1).
\eeq

For the axial-vector meson $K_{1A(B)}$, its %leading twist
twist-2 light-cone distribution amplitudes
can generally be expanded as the Gegenbauer polynomials~\cite{Yang:2007zt}:
\beq
 \phi_{K_{1A(B)}}(x)&=&\frac{f_{K_{1A(B)}}}{2\sqrt{2N_c}} 6 x  (1-x) \left[ a_{0K_{1A(B)}}^\parallel + 3
a_{1K_{1A(B)}}^\parallel\, (2x-1) + a_{2K_{1A(B)}}^\parallel\, \frac{3}{2} ( 5(2x-1)^2  - 1 )
\right]\;,\\
\phi_{K_{1A(B)}}^T(x)&=& \frac{f_{K_{1A(B)}}}{2\sqrt{2N_c}}6 x (1-x)
\left[ a_{0K_{1A(B)}}^\perp + 3 a_{1K_{1A(B)}}^\perp\, (2x-1) + a_{2K_{1A(B)}}^\perp\, \frac{3}{2} ( 5(2x-1)^2
- 1 ) \right] \;,
\eeq

For twist-3 light-cone distribution amplitudes, we use the following
form as in Ref.~\cite{Li:2009tx}:
\beq
\phi_{K_{1A(B)}}^s(x)&=&\frac{f_{K_{1A(B)}}}{4\sqrt{2N_c}} \frac{d}{dx}\Biggl[ 6x
(1-x) ( a_{0K_{1A(B)}}^\perp + a_{1K_{1A(B)}}^\perp (2x-1) )\Biggr]\;, \\  %\hspace{3mm}
\phi_{K_{1A(B)}}^t(x)&=&\frac{f_{K_{1A(B)}}}{2\sqrt{2N_c}}\Biggl[3a_{0K_{1A(B)}}^\perp (2x-1)^2+
\frac{3}{2}\,a_{1K_{1A(B)}}^\perp\,(2x-1) (3 (2x-1)^2-1)\Biggr],\\
\phi_{K_{1A(B)}}^v(x)&=&\frac{f_{K_{1A(B)}}}{2\sqrt{2N_c}}\Biggl[\frac{3}{4}
a_{0K_{1A(B)}}^\parallel
(1+(2x-1)^2) + \frac{3}{2}\, a_{1K_{1A(B)}}^\parallel\, (2x-1)^3\Biggr]\;,\\
\phi_{K_{1A(B)}}^a(x)&=&\frac{f_{K_{1A(B)}}}{8\sqrt{2N_c}}
\frac{d}{dx}\Biggl[6 x (1-x) ( a_{0K_{1A(B)}}^\parallel + a_{1K_{1A(B)}}^\parallel (2x-1))\Biggr]\;.
\eeq
where $f_{K_{1A(B)}}$ is the ``normalization" constant for both
longitudinally and transversely polarized mesons and the Gegenbauer
moments are quoted from Ref.~\cite{Yang:2007zt}
\begin{itemize}

\item  For $K_{1A}$ state,
\beq
a_0^{\parallel}&=& 1\;, \hspace{1.88cm}
a_1^{\parallel}=-0.30^{+0.00}_{-0.20}\;, \hspace{0.5cm}
a_2^{\parallel}=-0.05^{+0.03}_{-0.03}\;;
\non
a_0^{\perp}&=&0.27^{+0.03}_{-0.17}\;, \hspace{0.60cm}
a_1^{\perp}=-1.08^{+0.48}_{-0.48}\;, \hspace{0.46cm}
a_2^{\perp}=\;\;\; 0.02^{+0.21}_{-0.21}\;;
\eeq

\item For $K_{1B}$ state,
\beq
a_0^{\parallel}&=& -0.19^{+0.07}_{-0.07}\;,  \hspace{0.5cm}
a_1^{\parallel}=-1.95^{+0.45}_{-0.45}\;,   \hspace{0.5cm}
a_2^{\parallel}=\;\;\; 0.10^{+0.15}_{-0.19}\;;
\non
a_0^{\perp}&=&1\;, \hspace{1.71cm} \;\
a_1^{\perp}=\;\;\; 0.30^{+0.00}_{-0.33}\;, \hspace{0.44cm}
a_2^{\perp}= -0.02^{+0.22}_{-0.22}\;.
\eeq
\end{itemize}
where the values are taken at $\mu=1$ GeV.
Since $f_{K_{1A}}^\perp$ and $f_{K_{1B}}$ are G-parity-violating
quantities, their signs have to be
flipped from particle to antiparticle due to the G parity, for
example, $f_{K_{1B}^+} = -f_{K_{1B}^-}$. In the present work, the
G-parity violating parameters, e.g. $a_1^{\parallel,K_{1A}},
a_{0,2}^{\perp, K_{1A}}$, $a_1^{\perp, K_{1B}}$ and
$a_{0,2}^{\parallel, K_{1B}}$, are considered for mesons containing
a strange quark.

It is worth of mentioning that the $k_T$ dependence of the distribution amplitudes in the final
states has been neglected, since its contribution is very small as indicated in
Refs.~\cite{Li:1994cka,Li:1995jr,Li:1994iu}.
The underlying reason is that the contribution
from $k_T$ correlated with a soft dynamics is strongly suppressed by the Sudakov
effect through resummation for the wave function, which is dominated by a collinear dynamics.

\subsection{Perturbative calculations in pQCD approach} \label{ssec:pcalc}

For the considered $B \to \phi K_1$ decays induced by the %penguin dominant
$\bar b \to \bar s s \bar s$ transitions at the quark level,
the related weak effective
Hamiltonian $H_{{\rm eff}}$~\cite{Buchalla:1995vs} can be written as
\beq
H_{\rm eff}\, &=&\, {G_F\over\sqrt{2}}
\biggl\{ V^*_{ub}V_{us} [ C_1(\mu)O_1^{u}(\mu)
+C_2(\mu)O_2^{u}(\mu) ]
 - V^*_{tb}V_{ts} [ \sum_{i=3}^{10}C_i(\mu)O_i(\mu) ] \biggr\}+ {\rm H.c.}\;,
\label{eq:heff}
\eeq
with the Fermi constant $G_F=1.16639\times 10^{-5}{\rm
GeV}^{-2}$, Cabibbo-Kobayashi-Maskawa(CKM) matrix elements $V$,
and Wilson coefficients $C_i(\mu)$ at the renormalization scale
$\mu$. The local four-quark
operators $O_i(i=1,\cdots,10)$ are written as
\begin{enumerate}
\item[]{(1) current-current(tree) operators}
\begin{eqnarray}
{\renewcommand\arraystretch{1.5}
\begin{array}{ll}
\displaystyle
O_1^{u}\, =\,
(\bar{s}_\alpha u_\beta)_{V-A}(\bar{u}_\beta b_\alpha)_{V-A}\;,
& \displaystyle
O_2^{u}\, =\, (\bar{s}_\alpha u_\alpha)_{V-A}(\bar{u}_\beta b_\beta)_{V-A}\;;
\end{array}}
\label{eq:operators-1}
\end{eqnarray}

\item[]{(2) QCD penguin operators}
\begin{eqnarray}
{\renewcommand\arraystretch{1.5}
\begin{array}{ll}
\displaystyle
O_3\, =\, (\bar{s}_\alpha b_\alpha)_{V-A}\sum_{q'}(\bar{q}'_\beta q'_\beta)_{V-A}\;,
& \displaystyle
O_4\, =\, (\bar{s}_\alpha b_\beta)_{V-A}\sum_{q'}(\bar{q}'_\beta q'_\alpha)_{V-A}\;,
\\
\displaystyle
O_5\, =\, (\bar{s}_\alpha b_\alpha)_{V-A}\sum_{q'}(\bar{q}'_\beta q'_\beta)_{V+A}\;,
& \displaystyle
O_6\, =\, (\bar{s}_\alpha b_\beta)_{V-A}\sum_{q'}(\bar{q}'_\beta q'_\alpha)_{V+A}\;;
\end{array}}
\label{eq:operators-2}
\end{eqnarray}

\item[]{(3) electroweak penguin operators}
\begin{eqnarray}
{\renewcommand\arraystretch{1.5}
\begin{array}{ll}
\displaystyle
O_7\, =\,
\frac{3}{2}(\bar{s}_\alpha b_\alpha)_{V-A}\sum_{q'}e_{q'}(\bar{q}'_\beta q'_\beta)_{V+A}\;,
& \displaystyle
O_8\, =\,
\frac{3}{2}(\bar{s}_\alpha b_\beta)_{V-A}\sum_{q'}e_{q'}(\bar{q}'_\beta q'_\alpha)_{V+A}\;,
\\
\displaystyle
O_9\, =\,
\frac{3}{2}(\bar{s}_\alpha b_\alpha)_{V-A}\sum_{q'}e_{q'}(\bar{q}'_\beta q'_\beta)_{V-A}\;,
& \displaystyle
O_{10}\, =\,
\frac{3}{2}(\bar{s}_\alpha b_\beta)_{V-A}\sum_{q'}e_{q'}(\bar{q}'_\beta q'_\alpha)_{V-A}\;.
\end{array}}
\label{eq:operators-3}
\end{eqnarray}
\end{enumerate}
with the color indices $\alpha, \ \beta$ and the notations
$(\bar{q}'q')_{V\pm A} = \bar q' \gamma_\mu (1\pm \gamma_5)q'$.
The index $q'$ in the summation of the above operators runs
through $u,\;d,\;s$, $c$, and $b$.

%%%%=============================================================
\begin{figure}[!!htb]
\centering
\begin{tabular}{l}
\includegraphics[width=0.8\textwidth]{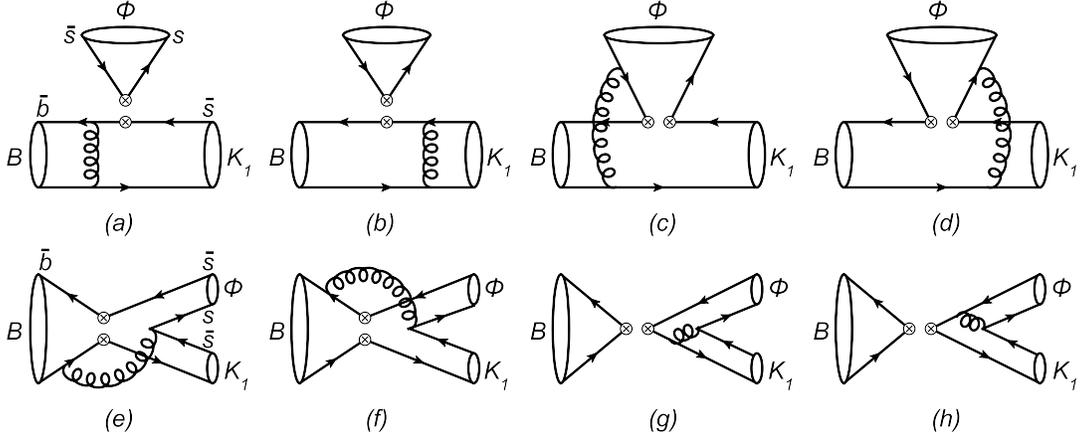}
\end{tabular}
\caption{Typical Feynman diagrams contributing to
the penguin-dominated $B \to \phi K_1$ decays in the pQCD approach
at leading order, in which $K_1$ stands for the axial-vector
$K_1(1270)$ and $K_1(1400)$, respectively.}
  \label{fig:fig1}
\end{figure}
%%%%==============================================================

From the effective Hamiltonian
(\ref{eq:heff}), there are eight types of diagrams contributing to the $B
\to \phi K_1$ decays in the pQCD approach at
leading order as illustrated in Fig.~\ref{fig:fig1}.
Analogous to the $B \to \phi K^*$ decays~\cite{Chen:2002pz},
we calculate the contributions arising from
various operators as shown in Eqs.~(\ref{eq:operators-1})-(\ref{eq:operators-3}).
Hereafter, for the sake of simplicity, we will use
%the decay amplitudes for
%the $B \to \phi K_{1A(B)}$
%decays can be obtained straightforwardly by replacing the
%variables of $K^*$ meson with those
%of $K_{1A(B)}$ states, apart from an overall minus sign arising from
%the definitions of the wave functions for vector $K^*$ meson
%and axial-vector $K_{1A(B)}$ mesons.
%Therefore, we here will
%not present the analytical expressions of factorization
%formulas for $B \to \phi K_{1A(B)}$ decays and the readers can refer
%to Ref.~\cite{Chen:2002pz} for them.
$F$ and $M$ to describe the factorizable
and non-factorizable amplitudes induced by the $(V-A)(V-A)$ operators,
$F^{P_1}$ and $M^{P_1}$ to describe the factorizable
and non-factorizable amplitudes arising from the $(V-A)(V+A)$ operators,
and $F^{P_2}$ and $M^{P_2}$ to describe the factorizable
and non-factorizable amplitudes coming from the $(S-P)(S+P)$ operators
that obtained by making Fierz transformation from the $(V-A)(V+A)$ operators,
respectively.
%And the subscripts ${\it fe}, {\it nfe}, {\it fa}$
%and ${\it nfa}$ are the abbreviations of factorizable emission, non-factorizable
%emission, factorizable annihilation, and non-factorizable annihilation,
%respectively.

For the factorizable emission($fe$) diagrams \ref{fig:fig1}(a) and \ref{fig:fig1}(b),
the corresponding Feynman amplitudes with one longitudinal polarization($L$) and
two transverse polarizations($N$ and $T$) can be read as follows,
\beq
F_{fe}^{L}&=&- 8\pi C_F m_{B}^2 \int_0^1
d x_{1} dx_{3}\, \int_{0}^{\infty} b_1 db_1 b_3 db_3\, \pb(x_1,b_1)
%\non && \times
\left\{ \left [(1+x_{3})\phi_{3}(x_{3}) +
r_{3}(1-2x_{3}) \right.\right. \non && \left. \left.\times
(\phi^{t}_{3}(x_3)+\phi^{s}_{3}(x_{3}))\right]E_{fe}(t_{a})
%\right. \non && \left.\times
h_{fe}(x_{1},x_{3},b_{1},b_{3})
+2r_{3}\phi^{s}_{3}(x_3)E_{fe}(t_{b})
h_{fe}(x_{3},x_{1},b_{3},b_{1}) \right\}\;,
\label{eq:abl}
\eeq
\beq
F_{fe}^{N}&=& -8\pi C_F m_{B}^2 \int_0^1 d x_{1}
dx_{3}\, \int_{0}^{\infty} b_1 db_1 b_3 db_3\, \phi_{B}(x_1,b_1)
%\non && \times
r_{2} \left\{ [\phi^{T}_{3}(x_3)
+2r_{3}\phi^{v}_{3}(x_3)+r_{3}x_{3}
\right. \non && \left. \times
(\phi^{v}_{3}(x_3)-\phi^{a}_{3}(x_3))]E_{fe}(t_{a})
%\right. \non &&
%\left.\times
h_{fe}(x_{1},x_{3},b_{1},b_{3})+r_{3}[\phi^{v}_{3}(x_3)+\phi^{a}_{3}(x_3)]
E_{fe}(t_{b}) h_{fe}(x_{3},x_{1},b_{3},b_{1})\right\}\;,
\label{eq:abn}
\eeq
\beq
 F_{fe}^{T}&=& -16\pi C_F m_{B}^2 \int_0^1 d x_{1} dx_{3}\,
\int_{0}^{\infty} b_1 db_1 b_3 db_3\, \phi_{B}(x_1,b_1)
%\non &&\times
r_{2}\left\{[\phi^{T}_{3}(x_3) +2r_{3}\phi^{a}_{3}(x_3)-r_{3}x_{3}
\right. \non &&
\left.\times
(\phi^{v}_{3}(x_3)-\phi^{a}_{3}(x_3))]E_{fe}(t_{a})  h_{fe}(x_{1},x_{3},b_{1},b_{3})
+r_{3}[\phi^{v}_{3}(x_3)+\phi^{a}_{3}(x_3)] E_{fe}(t_{b})h_{fe}(
x_{3},x_{1},b_{3},b_{1})\right\}\;; \label{eq:abt}
\eeq
where $\phi_{3}$ denotes the distribution amplitude of the
 axial-vector state $K_{1A}$ or $K_{1B}$ and
$C_F=4/3$ is a color factor. The hard functions $h_{i}$, the running hard scales $t_i$
and the convolution functions $E_{i}(t)$ can be referred to Ref.~\cite{Chen:2002pz}.

Since only the vector part of
$(V+A)$ current contributes to the vector meson production, $
\langle A |V-A|B\rangle \langle \phi |V+A | 0 \rangle = \langle A
|V-A |B \rangle \langle \phi |V-A|0 \rangle,$ that is
 \beq
 F_{fe}^{P_1}= F_{fe} \; \label{eq:abp1} .
 \eeq

For the non-factorizable emission($nfe$) diagrams \ref{fig:fig1}(c) and \ref{fig:fig1}(d),
the corresponding Feynman amplitudes are
\beq
 M_{nfe}^{L} &=& -\frac{16 \sqrt{6}}{3}\pi C_F m_{B}^2
\int_{0}^{1}d x_{1}d x_{2}\,d x_{3}\,\int_{0}^{\infty} b_1d b_1 b_2d
b_2\, \phi_{B}(x_1,b_1)\phi _{2 }(x_{2}) \left\{\left[
(1-x_{2}) \phi_{3} ( x_{3} )
\right. \right. \non && \left.  +r_{3}x_3
(\phi^{t}_{3}(x_3)-\phi^{s}_{3}(x_3)) \right]E_{nfe}(t_c)
h_{nfe}^{c}(x_1,x_2,x_3,b_1,b_2) -\left[
(x_{2}+x_{3}) \phi _{3} ( x_{3} )
\right.  \non && \left.\left.  -r_{3} x_3(\phi^{t}_{3}(x_3)+\phi^{s}_{3}(x_3))
\right] E_{nfe}(t_d)h_{nfe}^{d}(x_1,x_2,x_3,b_1,b_2) \right\}\;,
\label{eq:cdl}
\eeq
in which $\phi_{2}$ stands for the distribution amplitude of $\phi$ meson.
\beq
M_{nfe}^{N} &=& \frac{16 \sqrt{6}}{3}\pi C_F  m_{B}^2
\int_{0}^{1}d x_{1}d x_{2}\,d x_{3}\,\int_{0}^{\infty} b_1d b_1 b_2d
b_2\, \phi_{B}(x_1,b_1) r_{2}\left\{(1-x_2)
(\phi^{v}_{2}(x_2)+\phi^{a}_{2}(x_2) )
\right. \non && \left. \times
\phi^{T}_{3}(x_3) h_{nfe}^{c}(x_1,x_2,x_3,b_1,b_2) E_{nfe}(t_c)
+ \left[x_2(\phi^{v}_{2}(x_2) +\phi^a_{2}(x_2))\phi^{T}_{3}(x_3)
\right.\right. \non && \left. \left. -2r_{3}(x_2+x_3) (\phi^{v}_{2}(x_2) \phi^{v}_{3}(x_3)+
\phi^{a}_{2}(x_2) \phi^{a}_{3}(x_3) ) \right] E_{nfe}(t_d)
h_{nfe}^{d}(x_1,x_2,x_3,b_1,b_2) \right\}\;,
 \label{eq:cdn}
 \eeq
 \beq
 M_{nfe}^{T}&=& -\frac{32 \sqrt{6}}{3}\pi C_F m_{B}^2
\int_{0}^{1}d x_{1}d x_{2}\,d x_{3}\,\int_{0}^{\infty} b_1d b_1 b_2d
b_2\, \phi_{B}(x_1,b_1)   r_{2}\left\{
(1-x_2)(\phi^{v}_{2}(x_2)+\phi^{a}_{2}(x_2) )
\right. \non && \left. \times \phi^{T}_{3}(x_3)
h_{nfe}^{c}(x_1,x_2,x_3,b_1,b_2)
E_{nfe}(t_c)+   \left[
x_2(\phi^{v}_{2}(x_2)+\phi^a_{2}(x_2))\phi^{T}_{3}(x_3)
\right.\right. \non && \left.\left. -2r_{3}(x_2+x_3)
(\phi^{v}_{2}(x_2) \phi^{a}_{3}(x_3)+ \phi^{a}_{2}(x_2)
\phi^{v}_{3}(x_3) ) \right]E_{nfe}(t_d)
h_{nfe}^{d}(x_1,x_2,x_3,b_1,b_2) \right\}\; \label{eq:cdt}.
\eeq
 \beq
 M_{nfe}^{P_1,L} &=& -\frac{16 \sqrt{6}}{3}\pi C_F  m_{B}^2
\int_{0}^{1}d x_{1}d x_{2}\,d x_{3}\,\int_{0}^{\infty} b_1d b_1 b_2d
b_2\, \phi_{B}(x_1,b_1)  r_{2}
\left\{\left[(1-x_2)(\phi^{t}_{2}(x_2)+\phi^{s}_{2}(x_2))
\right. \right.\non && \times
\left. \phi_{3}(x_3)  -r_{3}(1-x_2) (\phi^{t}_{2}(x_2)
+\phi^{s}_{2}(x_2))(\phi^{t}_{3}(x_3)-\phi^{s}_{3}(x_3))-r_{3}
x_3(\phi^{t}_{2}(x_2) -\phi^{s}_{2}(x_2)) \right. \non && \left. \times
(\phi^{t}_{3}(x_3)+\phi^{s}_{3}(x_3)) \right]
 E_{nfe}(t_{c})h_{nfe}^{c}(x_{1},x_{2},x_{3},b_{1},b_{2})
+\left[ x_2 ( \phi^{t}_{2}(x_2)-\phi^{s}_{2}(x_2)) \phi_{3}(x_3)
\right.  \non &&  \left. \left.
-r_{3}x_2(\phi^{t}_{2}(x_2) -\phi^{s}_{2}(x_2)) (\phi^{t}_{3}(x_3)-\phi^{s}_{3}(x_3))
-r_{3}x_{3}(\phi^{t}_{2}(x_2)+\phi^{s}_{2}(x_2))
(\phi^{t}_{3}(x_3)+\phi^{s}_{3}(x_3)) \right] \right. \non &&  \left. \times
E_{nfe}(t_d)h_{nfe}^{d}(x_{1},x_{2},x_{3},b_{1},b_{2})
\right\}\;,\label{eq:cdl'}
\eeq
\beq
M_{nfe}^{P_1,N} &=&-\frac{16 \sqrt{6}}{3}\pi C_F m_{B}^2
\int_{0}^{1}d x_{1}d x_{2}\,d x_{3}\,\int_{0}^{\infty} b_1d b_1 b_2d
b_2\, \phi_{B}(x_1,b_1)
r_{3}x_{3} \phi^{T}_{2}(x_2) (\phi^{v}_{3}(x_3)-\phi^{a}_{3}(x_3))
 \non && \times
\left \{ E_{nfe}(t_{c})
h_{nfe}^{c}(x_{1},x_{2},x_{3},b_{1},b_{2})
+ E_{nfe}(t_{d}) h_{nfe}^{d}(x_{1},x_{2},x_{3},b_{1},b_{2}) \right\}\;,
\label{eq:cdn'}
\eeq
\beq
M_{nfe}^{P_1,T}&=& 2M_{nfe}^{P_1,N}\;,
\label{eq:cdt'}
\eeq
\beq
M_{nfe}^{P_2,L} &=& -\frac{16 \sqrt{6}}{3}\pi C_F  m_{B}^2
\int_{0}^{1}d x_{1}d x_{2}\,d x_{3}\,\int_{0}^{\infty} b_1d b_1 b_2d
b_2\, \phi_{B}(x_1,b_1)  \phi_{2}(x_2)\left\{
\left[(1-x_2+x_3)\phi_{3}(x_3)
\right.\right. \non &&  \left. \left.  - r_{3}x_3
(\phi^{t}_{3}(x_3)+\phi^{s}_{3}(x_3))\right]
E_{e}(t_{c}) h_{nfe}^{c}(x_{1},x_{2},x_{3},b_{1},b_{2})-
h_{nfe}^{d}(x_{1},x_{2},x_{3},b_{1},b_{2})E_{nfe}(t_{d})\right. \non && \times
\left[
x_2\phi_{3}(x_3)+r_{3}x_{3}(\phi^{t}_{3}(x_3)-\phi^{s}_{3}(x_3))
\right]\left.\right\}\;,
\eeq
\beq
M_{nfe}^{P_2,N} &=& \frac{16 \sqrt{6}}{3}\pi C_F m_{B}^2
\int_{0}^{1}d x_{1}d x_{2}\,d x_{3}\,\int_{0}^{\infty} b_1d b_1 b_2d
b_2\, \phi_{B}(x_1,b_1)  r_{2}
\left\{\left[(1-x_2)(\phi^{v}_{2}(x_2)-\phi^{a}_{2}(x_2))
\right. \right.\non && \left. \left. \times
\phi^{T}_{3}(x_3)-2r_{3}(1-x_2+x_3) (\phi^{v}_{2}(x_2)
\phi^{v}_{3}(x_3)-\phi^{a}_{2}(x_2)\phi^{a}_{3}(x_3))
\right]h_{nfe}^{c}(x_{1},x_{2},x_{3},b_{1},b_{2})
\right. \non &&\left. \times
E_{nfe}(t_{c})
 + x_2(\phi^{v}_{2}(x_2)-\phi^{a}_{2}(x_2))
\phi^{T}_{3}(x_3)  E_{nfe}(t_{d})
h_{nfe}^{d}(x_{1},x_{2},x_{3},b_{1},b_{2})\right\}\;,
\eeq
\beq
M_{nfe}^{P_2,T} &=& \frac{32 \sqrt{6}}{3}\pi C_F m_{B}^2
\int_{0}^{1}d x_{1}d x_{2}\,d x_{3}\,\int_{0}^{\infty} b_1d b_1 b_2d
b_2\, \phi_{B}(x_1,b_1) r_{2}
\left\{\left[(1-x_2)(\phi^{v}_{2}(x_2)-\phi^{a}_{2}(x_2))
\right. \right. \non && \left. \left.  \times
\phi^{T}_{3}(x_3)-2r_{3}(1-x_2+x_3) (\phi^{v}_{2}(x_2)
\phi^{a}_{3}(x_3)-\phi^{a}_{2}(x_2)\phi^{v}_{3}(x_3))
\right] h_{nfe}^{c}(x_{1},x_{2},x_{3},b_{1},b_{2})
\right. \non &&\left. \times
E_{nfe}(t_{c})+ x_2(\phi^{v}_{2}(x_2)-\phi^{a}_{2}(x_2))
\phi^{T}_{3}(x_3) E_{nfe}(t_{d})
h_{nfe}^{d}(x_{1},x_{2},x_{3},b_{1},b_{2})\right\}\;,
 \eeq

For the non-factorizable annihilation($nfa$) diagrams \ref{fig:fig1}(e)
and \ref{fig:fig1}(f), we have
\beq
M_{nfa}^{L} &=& -\frac{16\sqrt{6}}{3}\pi C_F m_{B}^2 \int_{0}^{1}d x_{1}d x_{2}\,d
x_{3}\,\int_{0}^{\infty} b_1d b_1 b_2d b_2\, \phi_{B}(x_1,b_1)
\left\{\left[ (1-x_3) \phi_{2}(x_2)\phi_{3}(x_3) \right.\right.\non
&& \left. \left. + r_{2}r_{3}
\left((1+x_2-x_3)(\phi^{s}_{2}(x_2)\phi^{s}_{3}(x_3)-\phi^{t}_{2}(x_2)
\phi^{t}_{3}(x_3))
-(1-x_2-x_3)(\phi^{s}_{2}(x_2)\phi^{t}_{3}(x_3)
\right.\right.\right.\non && \left.\left. \left.
-\phi^{t}_{2}(x_2)\phi^{s}_{3}(x_3) ) \right) \right]E_{nfa}(t_e)
 h_{nfa}^{e}(x_1,x_2,x_3,b_1,b_2)
-\left[ x_2\phi_{2}(x_2)\phi_{3}(x_3) + 2 r_{2}
r_{3}(\phi^{t}_{2}(x_2) \right.\right. \non &&
\left.\left. \times \phi^{t}_{3}(x_3)+\phi^{s}_{2}(x_2)\phi^{s}_{3}(x_3))-
r_{2}r_{3}(1+x_2-x_3)(\phi^{t}_{2}(x_2)\phi^{t}_{3}(x_3)
-\phi^{s}_{2}(x_2)\phi^{s}_{3}(x_3)) + r_{2} r_{3} \right.\right.\non
&&\left.\left. \times
(1-x_2-x_3)(\phi^{s}_{2}(x_2)\phi^{t}_{3}(x_3)
-\phi^{t}_{2}(x_2)\phi^{s}_{3}(x_3))\right] E_{nfa}(t_f)
 h_{nfa}^{f}(x_1,x_2,x_3,b_1,b_2) \right\}\;,
\label{eq:efl}
\eeq
\beq
M_{nfa}^{N} &=& \frac{32 \sqrt{6}}{3}\pi C_F m_{B}^2
\int_{0}^{1}d x_{1}d x_{2}\,d x_{3}\,\int_{0}^{\infty} b_1d b_1 b_2d
b_2\, \phi_{B}(x_1,b_1) r_{2} r_{3} \non &&\times
\left[\phi^{v}_{2}(x_2)\phi^{v}_{3}(x_3)
+\phi^{a}_{2}(x_2)\phi^{a}_{3}(x_3)\right] E_{nfa}(t_f)
h_{nfa}^{f}(x_1,x_2,x_3,b_1,b_2) \;,
 \label{eq:efn}
 \eeq
 \beq
M_{nfa}^{T}&=& \frac{64 \sqrt{6}}{3}\pi C_F m_{B}^2
\int_{0}^{1}d x_{1}d x_{2}\,d x_{3}\,\int_{0}^{\infty} b_1d b_1 b_2d
b_2\, \phi_{B}(x_1,b_1) r_{2} r_{3} \non &&\times
\left[\phi^{v}_{2}(x_2)\phi^{a}_{3}(x_3)
+\phi^{a}_{2}(x_2)\phi^{v}_{3}(x_3)\right] E_{nfa}(t_f)
h_{nfa}^{f}(x_1,x_2,x_3,b_1,b_2) \;
\label{eq:eft}.
\eeq
\beq
M_{nfa}^{P_1,L} &=& -\frac{16 \sqrt{6}}{3}\pi C_F  m_{B}^2
\int_{0}^{1}d x_{1}d x_{2}\,d x_{3}\,\int_{0}^{\infty} b_1d b_1 b_2d
b_2\, \phi_{B}(x_1,b_1) \left\{ \left[ r_{3}(1-x_3)
(\phi^{s}_{3}(x_3)- \phi^{t}_{3}(x_3)) \right. \right.
\non && \left. \left. \times
\phi_{2}(x_2)+r_{2}x_{2}(\phi^{t}_{2}(x_2)
+\phi^{s}_{2}(x_2))\phi_{3}(x_3)\right]
 E_{nfa}(t_{e})h_{nfa}^{e}(x_{1},x_{2},x_{3},b_{1},b_{2})-\left[ r_{2}(2-x_2)
\phi_{3}(x_3)
 \right.\right.\non && \left. \left. \times
 (\phi^{t}_{2}(x_2)+\phi^{s}_{2}(x_2))- r_{3}(1+x_3)\phi_{2}(x_2)
(\phi^{s}_{3}(x_3)-\phi^{t}_{3}(x_3))\right]
E_{nfa}(t_f)h_{nfa}^{f}(x_{1},x_{2},x_{3},b_{1},b_{2})
\right\}\;,\label{eq:efl'}
\eeq
\beq
M_{nfa}^{P_1,N} &=&-\frac{16 \sqrt{6}}{3}\pi C_F  m_{B}^2
\int_{0}^{1}d x_{1}d x_{2}\,d x_{3}\,\int_{0}^{\infty} b_1d b_1 b_2d
b_2\, \phi_{B}(x_1,b_1)
\left\{\left[r_{2}x_2(\phi^{v}_{2}(x_2) +\phi^{a}_{2}(x_2))
\phi^T_{3}(x_3)
\right.\right. \non && \left. \left.
- r_{3} (1-x_3) \phi^T_{2}(x_2)
(\phi^{a}_{3}(x_3)-\phi^{v}_{3}(x_3))\right]
 E_{nfa}(t_{e})h_{nfa}^{e}(x_{1},x_{2},x_{3},b_{1},b_{2})+\left[r_{2}(2-x_2)
 \phi^{T}_{3}(x_3)
 \right.\right.\non && \left. \left.\times
 (\phi^{v}_{2}(x_2) +\phi^{a}_{2}(x_2))  -r_{3}(1+x_3)\phi^{T}_{2}(x_2)
(\phi^{a}_{3}(x_3)-\phi^{v}_{3}(x_3)) \right]
E_{nfa}(t_f)h_{nfa}^{f}(x_{1},x_{2},x_{3},b_{1},b_{2})
\right\}\;,\label{eq:efn'}
\eeq
\beq
M_{nfa}^{P_1,T} &=& 2 M_{nfa}^{P_1,N}
\;,\label{eq:eft'}
 \eeq

For the factorizable annihilation($fa$) diagrams \ref{fig:fig1}(g) and \ref{fig:fig1}(h),
the contributions are
\beq
F_{fa}^{L}&=&- 8\pi C_F m_{B}^2 \int_0^1 d x_{2} dx_{3}\,
 \int_{0}^{\infty} b_2
db_2 b_3 db_3\, \left\{ \left [ x_{2}
\phi_{2}(x_{2})\phi_{3}(x_{3}) +2r_{3}r_{3}\phi^{s}_{3}(x_{3})
((1+x_{2}) \phi^{s}_{2}(x_2)
 \right. \right.\non && \left.\left. - (1-x_{2}) \phi^{t}_{2}(x_2))\right] E_{fa}(t_{g})
h_{fa}(x_{2},1-x_{3},b_{2},b_{3}) -
\left[(1-x_{3})\phi_{2}(x_2)\phi_{3}(x_3) + 2 r_{2} r_{3}
\phi^{s}_{2}(x_2) \right.\right. \non && \left. \left.  \times
(x_{3}\phi^{t}_{3}(x_3)+(2-x_{3})\phi^{s}_{3}(x_3))\right]E_{fa}(t_{h})
h_{fa}(1-x_{3},x_{2},b_{3},b_{2}) \right\}\;,
\label{eq:ghl}
\eeq
\beq
F_{fa}^{N}&=& -8\pi C_F m_{B}^2 \int_0^1 d x_{2}
dx_{3}\, \int_{0}^{\infty} b_2 db_2 b_3 db_3\,
r_{2}r_{3} \left\{ E_{fa}(t_{g})\left[(1+x_{2})(\phi^{v}_{2}(x_2)\phi^{v}_{3}(x_3)
+\phi^{a}_{2}(x_2)\phi^{a}_{3}(x_3))
\right. \right.  \non && \left.\left.
-(1-x_{2})(\phi^{v}_{2}(x_2)\phi^{a}_{3}(x_3)
+\phi^{a}_{2}(x_2)\phi^{v}_{3}(x_3))\right]
h_{fa}(x_{2},1-x_{3},b_{2},b_{3})-
\left[(2-x_{3})(\phi^{v}_{2}(x_2)\phi^{v}_{3}(x_3)\right.\right.\non
&& \left.\left.
+\phi^{a}_{2}(x_2)\phi^{a}_{3}(x_3))+x_3(\phi^{v}_{2}(x_2)\phi^{a}_{3}(x_3)
+\phi^{a}_{2}(x_2)\phi^{v}_{3}(x_3))\right] E_{fa}(t_{h})
h_{fa}( 1-x_{3},x_{2},b_{3},b_{2})\right\}\;,
\label{eq:ghn}
\eeq
\beq
F_{fa}^{T}&=& -16\pi C_F m_{B}^2 \int_0^1 d x_{2}
dx_{3}\, \int_{0}^{\infty} b_2 db_2 b_3 db_3\,
r_{2}r_{3}\left\{E_{fa}(t_{g})\left[(1+x_{2})(\phi^{v}_{2}(x_2)\phi^{a}_{3}(x_3)
+\phi^{a}_{2}(x_2)\phi^{v}_{3}(x_3))
\right.\right. \non &&
\left. \left.
-(1-x_{2})(\phi^{v}_{2}(x_2)\phi^{v}_{3}(x_3)
+\phi^{a}_{2}(x_2)\phi^{a}_{3}(x_3))\right]
h_{fa}(x_{2},1-x_{3},b_{2},b_{3})
+\left[x_{3}(\phi^{v}_{2}(x_2)\phi^{v}_{3}(x_3)\right.\right. \non
&& \left. \left.
+\phi^{a}_{2}(x_2)\phi^{a}_{3}(x_3))+(2-x_3)(\phi^{v}_{2}(x_2)\phi^{a}_{3}(x_3)
+\phi^{a}_{2}(x_2)\phi^{v}_{3}(x_3))\right] E_{fa}(t_{h})
 h_{fa}( 1-x_{3},x_{2},b_{3},b_{2})\right\}\;;
\label{eq:ght}
\eeq
\beq
F_{fa}^{P_2,L}&=&- 16\pi C_F m_{B}^2
\int_0^1 d x_{2} dx_{3}\, \int_{0}^{\infty}  b_2
db_2 b_3 db_3\, \left\{ \left[2 r_{3}
\phi_{2}(x_2)\phi^{s}_{3}(x_3)- r_{2} {x_2} (\phi^{t}_{2}(x_2) -
\phi^{s}_{2}(x_2) )
 \right.\right. \non && \left.\left.\times
\phi_{3}(x_3) \right]h_{fa}(x_{2},1-x_{3},b_{2},b_{3}) E_{fa}(t_{g})
+ \left[2 r_{2} \phi^{s}_{2}(x_2) \phi_{3}(x_3)+ r_{3} (1-x_3)
\phi_{2}(x_2)
\right.\right.\non && \left.\left. \times
(\phi^{t}_{3}(x_3)+\phi^{s}_{3}(x_3))\right]
E_{fa}(t_{h}) h_{fa}(1-x_{3},x_{2},b_{3},b_{2})
\right\}\;, \label{eq:ghl'}
\eeq
\beq
F_{fa}^{P_2,N}&=& -16\pi C_F
m_{B}^2 \int_0^1 d x_{2} dx_{3}\, \int_{0}^{\infty} b_2 db_2 b_3
db_3\,
\left\{r_{3}\phi^{T}_{2}(x_2)(\phi^{a}_{3}(x_3) -\phi^v_{3}(x_{3}))
 h_{fa}(x_{2},1-x_{3},b_{2},b_{3})
\right. \non && \left. \times
E_{fa}(t_{g})
+ r_{2}(\phi^{v}_{2}(x_2)
+\phi^a_{2}(x_{2}))\phi^{T}_{3}(x_3) E_{fa}(t_{h}) h_{fa}(
1-x_{3},x_{2},b_{3},b_{2})\right\}\;,
\label{eq:ghn'}
\eeq
\beq
F_{fa}^{P_2,T}&=& 2 F_{fa}^{P_2,N};
\label{eq:ght'}
\eeq

Before we put the things together to write down the decay amplitudes
for the considered $B \to \phi K_1$ modes,
it is essential to give a brief discussion about the
"$K_{1}(1270)-K_{1}(1400)$" mixing. The physical mass eigenstates $K_1(1270)$ and
$K_1(1400)$ are believed to be the mixtures of the $K_{1A}(^3P_1)$ and $K_{1B}(^1P_1)$ states with the
mixing angle $\theta_{K_1}$ due to the mass difference of the strange and
non-strange light quarks.
Following the common convention, their relations can be
written as~\cite{Beringer:1900zz}
\beq
\left(
\begin{array}{c} |K_1(1270)\rangle \\ |K_1(1400)\rangle \\ \end{array} \right ) &=&
  \left( \begin{array}{cc}
\sin{\theta_{K_1}} & \hspace{0.28cm} \cos{\theta_{K_1}} \\
 \cos{\theta_{K_1}} & -\sin\theta_{K_1} \end{array} \right )
 \left( \begin{array}{c}  |K_{1A}\rangle\\ |K_{1B} \rangle \\ \end{array} \right )\;,
 \label{eq:mixture}
\eeq
There exist several estimations on the mixing
angle $\theta_{K_1}$ in the literature~\cite{Suzuki:1993yc,Burakovsky:1997dd,Cheng:2003bn,
Hatanaka:2008xj,Cheng:2011pb,Divotgey:2013jba}.
Various phenomenological studies indicate that the $K_{1A}-K_{1B}$ mixing angle
$\theta_{K_1}$ is around either $33^\circ$ or $58^\circ$ but with a twofold ambiguity.
The sign ambiguity for $\theta_{K_1}$ is due to the fact that
one can add arbitrary phases to $|K_{1A}\rangle$ and
$|K_{1B}\rangle$.
As discussed in Ref.~\cite{Cheng:2011pb} and many early publications,
the sign ambiguity of $\theta_{K_1}$ can be removed by fixing the relative
sign of the decay constants of $|K_{1A}\rangle$ and $|K_{1B}\rangle$.
We shall choose the convention of decay constants in such a way
that $\theta_{K_1}$ is always positive.
It is noted that the sign of the mixing angle $\theta_{K_1}$
is positive for the mixing of particle states $K_{1A}$ and
$K_{1B}$ in this work, which corresponds to the negative
sign of $\theta_{K_1}$ between the mixing of antiparticle
states $\bar{K}_{1A}$ and $\bar{K}_{1B}$ in the
literature~\cite{Hatanaka:2008xj}.
The underlying reason is that, as discussed in Ref.~\cite{Blundell:1996as},  the spin-orbit portion
$<H_{q\bar q}^{SO-}>$
in the constituent quark model Hamiltonian causes the
mixing between $K_{1A}(^3P_1)$ and $K_{1B}(^1P_1)$ states, changes the sign when the antiquark instead of the quark is the heavier strange, then further leads to a mixing
angle of opposite sign when the $^3P_1-^1P_1$ Hamiltonian
is diagonalized.  In other words, if we have a angle of $-33^\circ$ for the mixing of
antiparticles $K_1(1270)^-$ and $K_1(1400)^-$, we must use $+33^\circ$ for that of the particles
$K_1(1270)^+$ and $K_1(1400)^+$.
For the value of mixing angle $\theta_{K_1}$,  we shall adopt both $33^\circ$ and $58^\circ$
in the numerical evaluations, which is because almost no any precise measurements on $\theta_{K_1}$
exist to date and one can identify the more favored value of the mixing angle in the
relevant $B$ meson decays, though Refs.~\cite{Hatanaka:2008xj,Cheng:2011pb,Cheng:2013cwa}
suggested that
the smaller angle $\theta_{K_1} \sim 33^\circ$ is much more favored than $58^\circ$.
%In fact, we have checked that, when $57^\circ$ is employed, the consistency with the available
%data for $B^+ \to \phi K_1(1270)^+$ decay deteriorates significantly. For example, in terms of
%the central values, the branching ratio will increase about 70\% while the longitudinal
%polarization fraction will decrease around 75\%.

Thus, by combining various contributions from different diagrams as
presented in Eqs.~(\ref{eq:abl})-(\ref{eq:ght'}) and the mixing pattern in Eq.~(\ref{eq:mixture}),
the total decay amplitudes for the penguin dominated $B \to \phi  K_1(1270)$ can be written as
\beq
{\cal M}_h(B^+ \to \phi K_1(1270)^+) &=& A_h(B^+ \to \phi
K_{1A}^+){\rm{\sin}}\theta_{K_1}  +  A_h(B^+ \to \phi
K_{1B}^+){\rm {\cos}}\theta_{K_1} \non &=& -\lambda_t f_{\phi}
\left( a_3 + a_4 + a_5 -\frac{1}{2}( a_7 + a_9+ a_{10} ) \right) (
F_{fe;K_{1A}}^{h}{\rm{\sin}}\theta_{K_1}+
F_{fe;K_{1B}}^{h}{\rm {\cos}}\theta_{K_1} )  \non & &- \lambda_t\biggl\{
( M_{nfe;K_{1A}}^{h}{\rm{\sin}}\theta_{K_1} +
M_{nfe;K_{1B}}^{h}{\rm {\cos}}\theta_{K_1} )
(C_3+C_4-\frac{1}{2}(C_9+C_{10}))\non && + (C_5-\frac{1}{2}C_7) (
M_{nfe;K_{1A}}^{P_1;h}{\rm{\sin}}\theta_{K_1}  +
M_{nfe;K_{1B}}^{P_1;h}{\rm
{\cos}}\theta_{K_1})+(C_6-\frac{1}{2}C_8)(
M_{nfe;K_{1A}}^{P_2;h}\non && \cdot {\rm{\sin}}\theta_{K_1} + {\rm
{\cos}}\theta_{K_1}  M_{nfe;K_{1B}}^{P_2;h}) \biggr\}
+\lambda_u C_1 (M_{nfa;K_{1A}}^{h}{\rm{\sin}}\theta_{K_1} +
M_{nfa;K_{1B}}^{h}{\rm {\cos}}\theta_{K_1})\non &&-\lambda_t
\biggl\{(C_3+C_9)(M_{nfa;K_{1A}}^{h}{\rm{\sin}}\theta_{K_1} +
M_{nfa;K_{1B}}^{h}{\rm {\cos}}\theta_{K_1})+(C_5+C_7)\non &&
\cdot (M_{nfa;K_{1A}}^{P_1;h}{\rm{\sin}}\theta_{K_1} +
M_{nfa;K_{1B}}^{P_1;h}{\rm {\cos}}\theta_{K_1})\biggr\}+\lambda_u
a_1(F_{fa;K_{1A}}^{h}{\rm{\sin}}\theta_{K_1} +
F_{fa;K_{1B}}^{h}\non &&
\cdot {\rm {\cos}}\theta_{K_1})f_B-\lambda_t
\biggl\{(a_4+a_{10})(F_{fa;K_{1A}}^{h}{\rm{\sin}}\theta_{K_1} +
F_{fa;K_{1B}}^{h} {\rm
{\cos}}\theta_{K_1})+(a_6+a_8)\non && \cdot (F_{fa;K_{1A}}^{P_2;h}{\rm{\sin}}\theta_{K_1}
 + F_{fa;K_{1B}}^{P_2;h}{\rm {\cos}}\theta_{K_1})\biggr\}f_B
\label{eq:mp};
\eeq
\beq
{\cal M}_h(B^0 \to \phi K_1(1270)^0) &=& A_h(B^0 \to
\phi K_{1A}^0){\rm{\sin}}\theta_{K_1}  +  A_h(B^0 \to \phi
K_{1B}^0){\rm {\cos}}\theta_{K_1} \non &=& -\lambda_t f_{\phi}
\left( a_3 + a_4 + a_5 -\frac{1}{2}( a_7 + a_9+ a_{10} ) \right)  (
F_{fe;K_{1A}}^{h}{\rm{\sin}}\theta_{K_1} +
F_{fe;K_{1B}}^{h}{\rm {\cos}}\theta_{K_1} ) \non & &- \lambda_t\biggl\{
( M_{nfe;K_{1A}}^{h}{\rm{\sin}}\theta_{K_1} +
M_{nfe;K_{1B}}^{h}{\rm {\cos}}\theta_{K_1} )
(C_3+C_4-\frac{1}{2}(C_9+C_{10}))\non &&+ (C_5-\frac{1}{2}C_7) (
M_{nfe;K_{1A}}^{P_1;h}{\rm{\sin}}\theta_{K_1}  +
M_{nfe;K_{1B}}^{P_1;h}{\rm {\cos}}\theta_{K_1})+(C_6-\frac{1}{2}C_8)
( M_{nfe;K_{1A}}^{P_2;h}\non && \cdot {\rm{\sin}}\theta_{K_1} + {\rm
{\cos}}\theta_{K_1} M_{nfe;K_{1B}}^{P_2;h}) \biggr\}
-\lambda_t
\biggl\{(C_3-\frac{1}{2}C_9)(M_{nfa;K_{1A}}^{h}{\rm{\sin}}\theta_{K_1}
+ M_{nfa;K_{1B}}^{h}\non && \cdot{\rm
{\cos}}\theta_{K_1})+(C_5-\frac{1}{2}C_7)
(M_{nfa;K_{1A}}^{P_1;h}{\rm{\sin}}\theta_{K_1} +
M_{nfa;K_{1B}}^{P_1;h}{\rm {\cos}}\theta_{K_1})\biggr\}-\lambda_tf_B
\biggl\{(a_4-\frac{1}{2}a_{10})\non && \cdot
(F_{fa;K_{1A}}^{h}{\rm{\sin}}\theta_{K_1}
+ F_{fa;K_{1B}}^{h} {\rm {\cos}}\theta_{K_1})%\non &&
+(a_6-\frac{1}{2}a_8)(F_{fa;K_{1A}}^{P_2;h}{\rm{\sin}}\theta_{K_1}
 + F_{fa;K_{1B}}^{P_2;h}{\rm {\cos}}\theta_{K_1})\biggr\}
\label{eq:m0};
\eeq
where $\lambda_u=V^*_{ub} V_{us}, \lambda_t=
V^*_{tb} V_{ts}$ and ${\cal M}_h$ in the above equations
denotes the different helicity
amplitudes ${\cal M}_L$, ${\cal
M}_N$, and ${\cal M}_T$, respectively.
And $a_i$ is the standard combination
of the Wilson coefficients $C_i$ defined as follows:
\beq
a_1 &=& C_2 + \frac{C_1}{3};
\quad a_i= C_i+C_{i\pm 1}/3,\quad  i=3-10.
\eeq
where $C_2 \sim 1$ is
the largest one among all Wilson coefficients and the upper (lower)
sign applies, when $i$ is odd (even).
When we make the replacements with ${\rm{\sin}}\theta_{K_1}\to
{\rm{\cos}}\theta_{K_1}, {\rm{\cos}}\theta_{K_1}\to
-{\rm{\sin}}\theta_{K_1}$ in Eqs.(\ref{eq:mp}) and (\ref{eq:m0}),
respectively, the total decay amplitudes for the $B \to \phi
K_1(1400)$ decays can be obtained straightforwardly.

\section{Numerical Results and Discussions} \label{sec:randd}

In this section, we will present the pQCD predictions on the
CP-averaged branching ratios, the polarization fractions, the relative
phases and the CP-violating asymmetries
for those four $B \to \phi K_1$ decay modes.
In numerical calculations, central values of the input parameters will be
used implicitly unless otherwise stated. The relevant QCD scale~({\rm GeV}), masses~({\rm GeV}),
and $B$ meson lifetime({\rm ps}) are the following
~\cite{Keum:2000ph,Keum:2000wi,Lu:2000em,Yang:2007zt,Beringer:1900zz}
\beq
 \Lambda_{\overline{\mathrm{MS}}}^{(f=4)} &=& 0.250\; , \quad m_W = 80.41\;,
 \quad  m_{B}= 5.2794\;,  \quad  m_b = 4.8 \;,\quad m_{\phi}= 1.02\;; \non
  f_{K_{1A}}&=& 0.250\;, \quad f_{K_{1B}} = 0.190\;,
\quad m_{K_{1A}}= 1.32\;,
\quad m_{K_{1B}}= 1.34\;, \non
  \tau_{B^+} &=& 1.641\;,  \quad \tau_{B^0}= 1.519\; .
\label{eq:mass}
\eeq

For the CKM matrix elements, we adopt the Wolfenstein
parametrization and the updated parameters $A=0.811$,
 $\lambda=0.22535$, $\bar{\rho}=0.131^{+0.026}_{-0.013}$, and $\bar{\eta}=0.345^{+0.013}_{-0.014}$~\cite{Beringer:1900zz}.

\subsection{CP-averaged branching ratios}

For the considered $B \to \phi K_1$ decays, the decay rate can be written as
\beq
\Gamma =\frac{G_{F}^{2}|\bf{P_c}|}{16 \pi m^{2}_{B} }
\sum_{\sigma=L,T}{\cal M}^{(\sigma)\dagger }{\cal M^{(\sigma)}}\;
\label{dr1}
\eeq
where $|\bf{P_c}|\equiv |\bf{P_{2z}}|=|\bf{P_{3z}}|$ is the momentum of either of the
outgoing axial-vector meson or vector meson and ${\cal M^{(\sigma)}}$ can be found in Eqs.~(\ref{eq:mp}-\ref{eq:m0}).
Using the decay amplitudes obtained in last section, it is straightforward to calculate the CP-averaged
branching ratios with uncertainties for the considered decays in the pQCD approach.
%%%========================================================================================================
\begin{table}[hbt]
\caption{ Theoretical predictions on physical quantities of $B^+ \to \phi K_1(1270)^+$
decay obtained in the pQCD approach with the mixing angle $\theta_{K_1} \sim 33^\circ$ and
$58^\circ$, respectively.
For comparison, we also quote the available experimental measurements~\cite{Aubert:2008bc}
and the estimations in the framework of QCD factorization~\cite{Cheng:2008gxa}.}
\label{tab:phik127p}
 \begin{center}\vspace{-0.3cm}{\tiny
\begin{tabular}[t]{c|c||c|c||c|c||c}
\hline  \hline
   \multicolumn{2}{c||}{Decay Mode}   &  \multicolumn{5}{c}{$B^+ \to \phi K_1(1270)^+$}  \\
   \hline
 Parameter  & Definition & pQCD ($\theta_{K_1} \sim 33^\circ$) &   QCDF ($\theta_{K_1} \sim 37^\circ$)
 & pQCD ($\theta_{K_1} \sim 58^\circ$) &   QCDF ($\theta_{K_1} \sim 58^\circ$) &  Experiment\\
\hline \hline
  BR($10^{-6}$)        & $\Gamma/ \Gamma_{\rm total}$
  &$5.4^{+0.9+0.5+3.3+2.1}_{-0.5-0.5-2.3-1.2}$
  &$3.8^{+1.9+5.1}_{-1.5-3.1}$
  &$9.2^{+0.2+1.3+4.4+2.5}_{-0.2-1.2-3.4-1.9}$
  &$3.4^{+2.2+5.5}_{-1.5-2.8}$&
  $6.1 \pm 1.6 \pm 1.1$
 \\
 \hline \hline
%%S
 $f_L$      & $|{\cal A}_L|^2$
 &$0.47^{+0.11+0.08+0.28+0.01}_{-0.09-0.06-0.30-0.00}$
 &$0.67^{+0.33}_{-0.64}$
 &$0.11^{+0.00+0.00+0.11+0.01}_{-0.01-0.02-0.07-0.01}$
 &$0.31^{+0.69}_{-0.37}$&
 $0.46^{+0.12+0.06}_{-0.13-0.07}$
 \\
 $f_{||}$   & $|{\cal A}_{||}|^2$
 &$0.30^{+0.06+0.03+0.17+0.00}_{-0.06-0.03-0.19-0.00}$
 &$-$
 &$0.45^{+0.01+0.00+0.06+0.00}_{-0.01-0.02-0.11-0.00}$
 &$-$
 &$-$
  \\
 $f_{\perp}$& $|{\cal A}_\perp|^2$
 &$0.22^{+0.04+0.05+0.17+0.00}_{-0.04-0.03-0.14-0.00}$
 &$-$
 &$0.45^{+0.01+0.00+0.05+0.01}_{-0.01-0.01-0.08-0.02}$
 &$-$
 &$-$
 \\
 \hline \hline
 $\phi_{||}$(rad)& $\arg\frac{{\cal A}_{||}}{{\cal A}_L}$
 &$2.2^{+0.1+0.0+0.3+0.1}_{-0.1-0.1-0.3-0.1}$
 &$-$
 &$3.3^{+0.0+0.1+0.6+0.0}_{-0.0-0.1-0.9-0.1}$
 &$-$
 &$-$
 \\
 $\phi_{\perp}$(rad)& $\arg\frac{{\cal A}_{\perp}}{{\cal A}_L}$
 &$4.4^{+0.0+0.1+0.2+0.0}_{-1.4-0.1-2.4-1.4}$
 &$-$
 &$2.6^{+0.0+0.1+0.7+0.0}_{-0.0-0.1-1.0-0.0}$
 &$-$
 &$-$
  \\
  \hline \hline
 $\acp^{\rm dir}(10^{-2})$& $\frac{\overline{\Gamma}-\Gamma}{\overline{\Gamma}+\Gamma}$
 &$-0.7^{+0.5+0.5+3.3+1.2}_{-0.3-0.4-2.6-1.3}$
 &$-$
 &$-1.3^{+0.0+0.3+0.9+0.5}_{-0.1-0.2-0.5-0.6}$
 &$-$
 &$-15\pm 19 \pm 5$
 \\
 $\acp^{\rm dir}(L)$& $\frac{\bar{f}_L-f_L}{\bar{f}_L+f_L}$
 & $0.04^{+0.09}_{-0.07}$
 & $-$
 & $0.16^{+0.08}_{-0.10}$
 & $-$
 & $-$
 \\
 $\acp^{\rm dir}(||)$& $\frac{\bar{f}_{||}-f_{||}}{\bar{f}_{||}+f_{||}}$
 &$-0.33^{+0.12}_{-0.14}$
 &$-$
 &$-0.29^{+0.09}_{-0.08}$
 &$-$
 &$-$
 \\
 $\acp^{\rm dir}(\perp)$& $\frac{\bar{f}_\perp-f_\perp}{\bar{f}_\perp+f_\perp}$
 &$0.33^{+0.11}_{-0.09}$
 &$-$
 &$0.22^{+0.05}_{-0.05}$
 &$-$
 &$-$
 \\ \hline \hline
\end{tabular}}
\end{center}
\end{table}
%% ==========================================================================================================
%The theoretical predictions for the CP-averaged
%branching ratios of the considered penguin dominated $B \to \phi K_1$ decays in the
%pQCD approach can be read as follows,
The numerical results of the physical quantities are presented in Tables~\ref{tab:phik127p}-\ref{tab:phik1400},
in which the major errors are induced by the uncertainties
of the shape parameter $\omega_b = 0.40 \pm 0.04$~GeV
for the $B$ meson wave function, of the vector $\phi$ meson
decay constants $f_{\phi}=0.231 \pm 0.004$~GeV
and $f_{\phi}^T=0.200 \pm 0.010$~GeV
and the axial-vector $K_{1A}$
and $K_{1B}$ states decay constants $f_{K_{1A}}=0.250 \pm 0.013$~GeV
and $f_{K_{1B}}=0.190 \pm 0.010$~GeV, and
of the Gegenbauer moments $a_i^{A,B}(i=0,1,2)$
for the axial-vector $K_{1A}$ and $K_{1B}$ states
and $a_{2}$ for the vector $\phi$ meson in both longitudinal and transverse
polarizations, respectively.
Moreover, in this work, as displayed in the above mentioned Tables,
the higher order contributions are also simply investigated by exploring the variation of the hard scale $t_{\rm max}$, i.e., from $0.8t$ to $1.2t$
(not changing $1/b_i, i= 1,2,3$), in the hard kernel, which have been counted into one of the source of theoretical uncertainties(See the last term of errors in the related Tables).
Note that the variation of the
CKM parameters has tiny or almost no effects
to the physical observables of
these $B \to \phi K_1$ decays in the pQCD approach
and thus have been neglected in the relevant numerical results.

%%%========================================================================================================
\begin{table}[hbt]
\caption{ Same as Table~\ref{tab:phik127p} but of $B^+ \to \phi K_1(1400)^+$ decay.}
\label{tab:phik140p}
 \begin{center}\vspace{-0.3cm}{\tiny
\begin{tabular}[t]{c|c||c|c||c|c||c}
\hline  \hline
   \multicolumn{2}{c||}{Decay Mode}   &  \multicolumn{5}{c}{$B^+ \to \phi K_1(1400)^+$}  \\
   \hline
 Parameter  & Definition & pQCD ($\theta_{K_1} \sim 33^\circ$) &   QCDF ($\theta_{K_1} \sim 37^\circ$)
 & pQCD ($\theta_{K_1} \sim 58^\circ$) &   QCDF ($\theta_{K_1} \sim 58^\circ$) &  Experiment\\
\hline \hline
  BR($10^{-6}$)        & $\Gamma/ \Gamma_{\rm total}$
  &$25.1^{+8.5+2.1+10.7+4.4}_{-6.0-2.2-8.9-3.7}$
  &$11.1^{+8.5+41.1}_{-5.4-11.4}$
  &$21.4^{+9.3+2.0+7.9+3.9}_{-6.3-1.8-6.7-3.0}$
  &$11.3^{+7.5+40.2}_{-4.9-11.1}$&
  $<3.2(0.3\pm 1.6\pm 0.7)(90\%\ {\rm C.L.})$
 \\
 \hline \hline
%%S
 $f_L$      & $|{\cal A}_L|^2$
 &$0.57^{+0.06+0.03+0.12+0.02}_{-0.06-0.02-0.11-0.02}$
 &$0.45^{+0.13}_{-0.09}$
 &$0.74^{+0.04+0.02+0.10+0.02}_{-0.04-0.02-0.09-0.00}$
 &$0.57^{+0.32}_{-0.22}$&
 $-$
 \\
 $f_{||}$   & $|{\cal A}_{||}|^2$
 &$0.20^{+0.03+0.01+0.07+0.01}_{-0.03-0.02-0.06-0.01}$
 &$-$
 &$0.12^{+0.02+0.00+0.06+0.00}_{-0.02-0.00-0.05-0.01}$
 &$-$
 &$-$
  \\
 $f_{\perp}$& $|{\cal A}_\perp|^2$
 &$0.23^{+0.03+0.00+0.05+0.01}_{-0.03-0.02-0.05-0.01}$
 &$-$
 &$0.14^{+0.02+0.00+0.04+0.00}_{-0.02-0.02-0.05-0.01}$
 &$-$
 &$-$
 \\
 \hline \hline
 $\phi_{||}$(rad)& $\arg\frac{{\cal A}_{||}}{{\cal A}_L}$
 &$4.1^{+0.1+0.0+0.2+0.1}_{-0.1-0.0-0.2-0.0}$
 &$-$
 &$4.0^{+0.1+0.0+0.2+0.1}_{-0.1-0.0-0.3-0.0}$
 &$-$
 &$-$
 \\
 $\phi_{\perp}$(rad)& $\arg\frac{{\cal A}_{\perp}}{{\cal A}_L}$
 &$3.7^{+0.0+0.0+0.2+0.1}_{-0.0-0.0-0.2-0.1}$
 &$-$
 &$3.8^{+0.0+0.0+0.2+0.1}_{-0.0-0.0-0.2-0.1}$
 &$-$
 &$-$
  \\
  \hline \hline
 $\acp^{\rm dir}(10^{-2})$& $\frac{\overline{\Gamma}-\Gamma}{\overline{\Gamma}+\Gamma}$
 &$-1.5^{+0.3+0.1+1.1+0.3}_{-0.4-0.1-1.4-0.3}$
 &$-$
 &$-1.3^{+0.2+0.0+1.4+0.5}_{-0.4-0.2-1.6-0.5}$
 &$-$
 &$-$
 \\
 $\acp^{\rm dir}(L)$& $\frac{\bar{f}_L-f_L}{\bar{f}_L+f_L}$
 & $-0.01^{+0.02}_{-0.02}$
 & $-$
 & $-0.01^{+0.01}_{-0.04}$
 & $-$
 & $-$
 \\
 $\acp^{\rm dir}(||)$& $\frac{\bar{f}_{||}-f_{||}}{\bar{f}_{||}+f_{||}}$
 &$-0.17^{+0.05}_{-0.03}$
 &$-$
 &$-0.08^{+0.05}_{-0.06}$
 &$-$
 &$-$
 \\
 $\acp^{\rm dir}(\perp)$& $\frac{\bar{f}_\perp-f_\perp}{\bar{f}_\perp+f_\perp}$
 &$0.11^{+0.02}_{-0.04}$
 &$-$
 &$0.05^{+0.03}_{-0.06}$
 &$-$
 &$-$
 \\ \hline \hline
\end{tabular}}
\end{center}
\end{table}
%% ==========================================================================================================

Based on the theoretical branching ratios given at leading order in the pQCD approach,
some phenomenological remarks on the $B \to \phi K_1$ decays are in order:

\begin{itemize}

\item
From Table~\ref{tab:phik127p}, one can easily find that the CP-averaged branching ratios of
$B^+ \to \phi K_1(1270)^+$ decay are
\beq
Br(B^+ \to \phi K_1(1270)^+)_{\rm pQCD} &=&
\left\{ \begin{array}{ll}
5.4^{+4.0}_{-2.7} \times 10^{-6}\;\;\;\;\;\;\;\;{\rm \theta_{K_1} \sim 33^\circ}&  \vspace{0.1cm}\\
9.2^{+5.2}_{-4.1} \times 10^{-6}\;\;\;\;\;\;\;\;{\rm \theta_{K_1} \sim 58^\circ} &  \\ \end{array} \right. \;,
\eeq
where various errors arising from the input parameters have been added in quadrature.
It is observed that the former prediction with the smaller angle $33^\circ$ is more consistent with the available measurement~\cite{Aubert:2008bc},
 \beq
 Br(B^\pm \to \phi K_1(1270)^\pm)_{\rm Exp.} &=& (6.1 \pm 1.9) \times 10^{-6}\;,
 \eeq
where the systematic and statistical errors have also been added in quadrature,
although the latter prediction basically agrees with the theoretical values obtained in the
framework of QCD factorization and the preliminary data reported by BABAR Collaboration
within large errors.

\item
According to Table~\ref{tab:phik140p}, the CP-averaged branching ratios
of $B^+ \to \phi K_1(1400)^+$ decay with two different mixing angles can be read as
\beq
Br(B^+ \to \phi K_1(1400)^+)_{\rm pQCD} &=&
\left\{ \begin{array}{ll}
25.1^{+14.5}_{-11.6} \times 10^{-6}\;\;\;\;\;\;\;\;{\rm \theta_{K_1} \sim 33^\circ}&  \vspace{0.1cm}\\
21.4^{+13.0}_{-9.8} \times 10^{-6}\;\;\;\;\;\;\;\;{\rm \theta_{K_1} \sim 58^\circ} &  \\ \end{array} \right. \;,
\eeq
Here, we have added all the errors in quadrature.
Our theoretical predictions are in agreement with that derived in the QCDF approach within large
errors and also with the preliminary upper limit~\cite{Aubert:2008bc}
 \beq
 Br(B^\pm \to \phi K_1(1400)^\pm)_{\rm Exp.} &<& 3.2 (0.3\pm 1.6 \pm 0.7) \times 10^{-6}\;,
 \eeq
in 2$\sigma$ errors roughly. But,
it looks that, unfortunately, the central values
significantly exceed the upper limit placed by only BABAR Collaboration.
It will be very interesting and probably a challenge for the
theorists to further understand the QCD dynamics of
these two strange axial-vector mesons and the mixing between $K_{1A}$ and $K_{1B}$ states in depth
once the experiments at LHC and/or Super-B confirm the aforementioned much small upper limits of $Br(B^+ \to \phi K_1(1400)^+)$ in the near future.

\item
As can be seen in Tables~\ref{tab:phik1270} and \ref{tab:phik1400}, the CP-averaged branching ratios
of $B^0 \to \phi K_1^0$ decays with two different mixing angles are
also predicted in the pQCD approach,
\beq
Br(B^0 \to \phi K_1(1270)^0)_{\rm pQCD} &=&
\left\{ \begin{array}{ll}
5.1^{+3.5}_{-2.7} \times 10^{-6} \;\;\;\;\;\;\;\;{\rm \theta_{K_1} \sim 33^\circ}& \vspace{0.1cm} \\
9.2^{+5.5}_{-4.0} \times 10^{-6} \;\;\;\;\;\;\;\;{\rm \theta_{K_1} \sim 58^\circ} &  \\ \end{array} \right. \;, \\
Br(B^0 \to \phi K_1(1400)^0)_{\rm pQCD} &=& \left\{ \begin{array}{ll}
22.5^{+12.8}_{-10.3} \times 10^{-6}\;\;\;\;\;\;\;\;{\rm \theta_{K_1} \sim 33^\circ}& \vspace{0.1cm} \\
18.5^{+11.3}_{-8.8} \times 10^{-6}\;\;\;\;\;\;\;\;{\rm \theta_{K_1} \sim 58^\circ} &  \\ \end{array} \right. \;.
\eeq
which are consistent with the predictions in the QCDF approach
within large theoretical errors and will be tested in the running LHC and forthcoming Super-B
experiments.

\item
As discussed in Refs.~\cite{Yang:2007zt,Cheng:2008gxa},
the behavior of the axial-vector $^3P_1$ states is similar to that of
the vector mesons, which will consequently result in the branching ratio
of $B \to \phi K_{1A}$ analogous to that of $B \to \phi K^*$ decays in the pQCD approach.
However, from Tables~\ref{tab:phik127p}-\ref{tab:phik1400}, it can be clearly observed that
the predicted branching ratios of $B \to \phi K_1(1270)(B \to \phi K_1(1400))$ decays
in the pQCD approach are smaller(larger) than those of $B \to \phi K^*$ decays~\cite{Chen:2002pz},
which imply the destructive(constructive) effects between $B \to \phi K_{1A}$
and $B \to \phi K_{1B}$ decay amplitudes to $B \to \phi K_1(1270)(B \to \phi K_1(1400))$ decays.
In order to clarify this point more clearly, we present the decay amplitudes
of the $B^+ \to \phi K_{1A}^+$ and $B^+ \to \phi K_{1B}^+$ decays numerically for
every topology with three polarizations, which can be seen in
Table~\ref{tab:DA-pm}.

%%%========================================================================================================
\begin{table}[t]
\caption{ Same as Table~\ref{tab:phik127p} but of $B^0 \to \phi K_1(1270)^0$ decay.}
\label{tab:phik1270}
 \begin{center}\vspace{-0.3cm}{\tiny
\begin{tabular}[t]{c|c||c|c||c|c||c}
\hline  \hline
   \multicolumn{2}{c||}{Decay Mode}   &  \multicolumn{5}{c}{$B^0 \to \phi K_1(1270)^0$}  \\
   \hline
 Parameter  & Definition & pQCD ($\theta_{K_1} \sim 33^\circ$) &   QCDF ($\theta_{K_1} \sim 37^\circ$)
 & pQCD ($\theta_{K_1} \sim 58^\circ$) &   QCDF ($\theta_{K_1} \sim 58^\circ$) &  Experiment\\
\hline \hline
  BR($10^{-6}$)        & $\Gamma/ \Gamma_{\rm total}$
  &$5.1^{+0.7+0.4+2.9+1.8}_{-0.5-0.5-2.3-1.2}$
  &$3.6^{+1.7+4.8}_{-1.3-2.9}$
  &$9.2^{+0.2+1.3+4.7+2.5}_{-0.1-1.2-3.4-1.8}$
  &$3.2^{+2.1+5.2}_{-1.4-2.7}$&
  $-$
 \\
 \hline \hline
%%S
 $f_L$      & $|{\cal A}_L|^2$
 &$0.42^{+0.11+0.06+0.30+0.01}_{-0.09-0.06-0.28-0.00}$
 &$0.67^{+0.33}_{-0.64}$
 &$0.11^{+0.00+0.02+0.13+0.01}_{-0.00-0.01-0.06-0.00}$
 &$0.31^{+0.69}_{-0.31}$&
 $-$
 \\
 $f_{||}$   & $|{\cal A}_{||}|^2$
 &$0.35^{+0.05+0.02+0.14+0.00}_{-0.07-0.04-0.20-0.01}$
 &$-$
 &$0.45^{+0.01+0.00+0.07+0.01}_{-0.01-0.00-0.11-0.00}$
 &$-$
 &$-$
  \\
 $f_{\perp}$& $|{\cal A}_\perp|^2$
 &$0.24^{+0.03+0.03+0.13+0.00}_{-0.05-0.05-0.17-0.01}$
 &$-$
 &$0.44^{+0.01+0.00+0.05+0.01}_{-0.01-0.01-0.07-0.02}$
 &$-$
 &$-$
 \\
 \hline \hline
 $\phi_{||}$(rad)& $\arg\frac{{\cal A}_{||}}{{\cal A}_L}$
 &$2.3^{+0.1+0.1+0.4+0.1}_{-0.1-0.1-0.3-0.1}$
 &$-$
 &$3.4^{+0.0+0.1+0.5+0.0}_{-0.0-0.1-0.8-0.1}$
 &$-$
 &$-$
 \\
 $\phi_{\perp}$(rad)& $\arg\frac{{\cal A}_{\perp}}{{\cal A}_L}$
 &$4.4^{+0.2+0.1+0.5+0.2}_{-0.2-0.1-0.4-0.2}$
 &$-$
 &$2.7^{+0.0+0.1+0.6+0.0}_{-0.0-0.1-0.9-0.0}$
 &$-$
 &$-$
  \\
  \hline \hline
 $\acp^{\rm dir}(10^{-2})$& $\frac{\overline{\Gamma}-\Gamma}{\overline{\Gamma}+\Gamma}$
 &$0.0$
 &$-$
 &$0.0$
 &$-$
 &$-$
 \\
 $\acp^{\rm dir}(L)$& $\frac{\bar{f}_L-f_L}{\bar{f}_L+f_L}$
 & $0.0$
 & $-$
 & $0.0$
 & $-$
 & $-$
 \\
 $\acp^{\rm dir}(||)$& $\frac{\bar{f}_{||}-f_{||}}{\bar{f}_{||}+f_{||}}$
 &$0.0$
 &$-$
 &$0.0$
 &$-$
 &$-$
 \\
 $\acp^{\rm dir}(\perp)$& $\frac{\bar{f}_\perp-f_\perp}{\bar{f}_\perp+f_\perp}$
 &$0.0$
 &$-$
 &$0.0$
 &$-$
 &$-$
 \\ \hline \hline
\end{tabular}}
\end{center}
\end{table}
%% ==========================================================================================================

\item
As mentioned in the Introduction, up to now, the penguin-dominated $B \to \phi
K_1$ decays have been investigated with different approaches/methods~\cite{Calderon:2007nw,Chen:2005cx,Cheng:2008gxa}.
With the form factors of $B \to K_1$ transitions calculated in the improved
Isgur-Scora-Grinstein-Wise quark model, the authors got the branching ratios
of $B \to \phi K_1(1270)$ and $B \to \phi K_1(1400)$ decays with
two different mixing angles $32^\circ$ and $58^\circ$~\cite{Calderon:2007nw}
in the naive factorization approach.
However, the results of the former modes are too small($10^{-9} \sim 10^{-7}$) to be
comparable with the available measurements and that for the latter ones are
consistent with the preliminary upper limits.
Those branching ratios %predicted in the naive factorization approach
indicate the
the destructive(constructive) interferences between $B \to \phi K_{1A}$
and $B \to \phi K_{1B}$. With the $B \to K_{1A}$ and $B \to K_{1B}$ form factors
taken from light-front quark model and by neglecting the so-called "negligible" annihilation contributions, the authors obtained the ${\cal O}(10^{-5})$ and ${\cal O}(10^{-6})$
branching ratios in the generalized factorization approach for
$B \to \phi K_1(1270)$ and $B \to \phi K_1(1400)$ decays, respectively,
when the preferred effective color number $N_c^{eff}$ is $2$ or $3$,
which exhibit the constructive(destructive) contributions to $B \to \phi K_1(1270)
(B \to \phi K_1(1400))$ modes and
the contrary decay pattern to that given in Ref.~\cite{Calderon:2007nw}.

\item
Armed with the light-cone wave functions of
axial-vector mesons in QCD sum rule method,
Cheng and Yang studied the $B \to \phi K_1$ decays explicitly in the QCDF
approach~\cite{Cheng:2008gxa}. The predictions for the branching ratios
of the considered $B \to \phi K_1$ decays in QCDF are also presented in
Tables~\ref{tab:phik127p}-\ref{tab:phik1400}.
It is necessary to point out that the evaluations on these $B \to \phi K_1$ decays in QCDF
have used the weak annihilation parameters, which can be
sizable and important on polarizations~\cite{Chen:2005cx}, inferred from the
vector-vector $B \to \phi K^*$ decays. The QCDF predictions show
the similar interferences between $B \to \phi K_{1A}$ and $B \to \phi K_{1B}$
to that shown in the pQCD approach, which is more apparent in both predictions with
the smaller mixing angle $33^\circ$. Moreover, according to the CP-averaged
branching ratios of $B \to \phi K_1$ decays, the QCDF results show the weak dependence
of mixing angle, while the pQCD values exhibit the stronger(weaker) sensitivity to the
mixing angle in $B \to \phi K_1(1270)(B \to \phi K_1(1400))$ decays.
The underlying reason is that with the increasing
of the mixing angle $\theta_{K_1}$, the significantly destructive interferences on longitudinal
polarization and dramatically constructive effects(specifically, in the annihilation diagrams) on both transverse polarizations between
$B \to \phi K_{1A}$ and $B \to \phi K_{1B}$(See Table~\ref{tab:DA-pm}) result
in the large branching ratios but small
longitudinal polarization fraction in $B \to \phi K_1(1270)$ decays.
%%%========================================================================================================
\begin{table}[hbt]
\caption{ Same as Table~\ref{tab:phik127p} but of $B^0 \to \phi K_1(1400)^0$ decay.}
\label{tab:phik1400}
 \begin{center}\vspace{-0.3cm}{\tiny
\begin{tabular}[t]{c|c||c|c||c|c||c}
\hline  \hline
   \multicolumn{2}{c||}{Decay Mode}   &  \multicolumn{5}{c}{$B^0 \to \phi K_1(1400)^0$}  \\
   \hline
 Parameter  & Definition & pQCD ($\theta_{K_1} \sim 33^\circ$) &   QCDF ($\theta_{K_1} \sim 37^\circ$)
 & pQCD ($\theta_{K_1} \sim 58^\circ$) &   QCDF ($\theta_{K_1} \sim 58^\circ$) &  Experiment\\
\hline \hline
  BR($10^{-6}$)        & $\Gamma/ \Gamma_{\rm total}$
  &$22.5^{+7.7+2.0+9.2+4.0}_{-5.3-2.1-7.9-3.3}$
  &$10.4^{+7.9+38.3}_{-5.1-10.4}$
  &$18.5^{+8.2+1.8+6.7+3.4}_{-5.6-1.8-5.9-2.7}$
  &$10.7^{+7.1+37.7}_{-4.6-10.4}$&
  $-$
 \\
 \hline \hline
%%S
 $f_L$      & $|{\cal A}_L|^2$
 &$0.55^{+0.06+0.02+0.10+0.00}_{-0.07-0.03-0.13-0.00}$
 &$0.46^{+0.26}_{-0.02}$
 &$0.72^{+0.04+0.02+0.11+0.02}_{-0.05-0.02-0.10-0.01}$
 &$0.57^{+0.31}_{-0.22}$&
 $-$
 \\
 $f_{||}$   & $|{\cal A}_{||}|^2$
 &$0.21^{+0.03+0.01+0.06+0.00}_{-0.03-0.02-0.07-0.00}$
 &$-$
 &$0.12^{+0.03+0.02+0.07+0.01}_{-0.01-0.00-0.04-0.00}$
 &$-$
 &$-$
  \\
 $f_{\perp}$& $|{\cal A}_\perp|^2$
 &$0.25^{+0.03+0.00+0.05+0.00}_{-0.04-0.02-0.06-0.01}$
 &$-$
 &$0.15^{+0.03+0.02+0.06+0.01}_{-0.02-0.00-0.04-0.01}$
 &$-$
 &$-$
 \\
 \hline \hline
 $\phi_{||}$(rad)& $\arg\frac{{\cal A}_{||}}{{\cal A}_L}$
 &$4.2^{+0.1+0.0+0.2+0.1}_{-0.1-0.0-0.2-0.0}$
 &$-$
 &$4.1^{+0.1+0.0+0.2+0.1}_{-0.1-0.0-0.3-0.0}$
 &$-$
 &$-$
 \\
 $\phi_{\perp}$(rad)& $\arg\frac{{\cal A}_{\perp}}{{\cal A}_L}$
 &$3.7^{+0.0+0.0+0.2+0.1}_{-0.0-0.0-0.2-0.1}$
 &$-$
 &$3.8^{+0.1+0.0+0.3+0.1}_{-0.0-0.0-0.3-0.1}$
 &$-$
 &$-$
  \\
  \hline \hline
 $\acp^{\rm dir}(10^{-2})$& $\frac{\overline{\Gamma}-\Gamma}{\overline{\Gamma}+\Gamma}$
 &$0.0$
 &$-$
 &$0.0$
 &$-$
 &$-$
 \\
 $\acp^{\rm dir}(L)$& $\frac{\bar{f}_L-f_L}{\bar{f}_L+f_L}$
 & $0.0$
 & $-$
 & $0.0$
 & $-$
 & $-$
 \\
 $\acp^{\rm dir}(||)$& $\frac{\bar{f}_{||}-f_{||}}{\bar{f}_{||}+f_{||}}$
 &$0.0$
 &$-$
 &$0.0$
 &$-$
 &$-$
 \\
 $\acp^{\rm dir}(\perp)$& $\frac{\bar{f}_\perp-f_\perp}{\bar{f}_\perp+f_\perp}$
 &$0.0$
 &$-$
 &$0.0$
 &$-$
 &$-$
 \\ \hline \hline
\end{tabular}}
\end{center}
\end{table}
%% ==========================================================================================================

\item
In view of the large theoretical errors from the hadronic parameters
in the pQCD predictions, we define the interesting ratios
as follows,
 \beq
 \frac{Br(B^+ \to \phi K_1(1270)^+)}{Br(B^+ \to \phi K_1(1400)^+)}
 &=&
\left\{ \begin{array}{ll}
 0.22^{+0.20}_{-0.15} \;\;\;\;\;\;\;\;{\rm \theta_{K_1} \sim 33^\circ}& \vspace{0.1cm} \\
0.43^{+0.36}_{-0.27} \;\;\;\;\;\;\;\;{\rm \theta_{K_1} \sim 58^\circ} &  \\ \end{array} \right. \;,
\eeq
\beq
\frac{Br(B^0 \to \phi K_1(1270)^0)}{Br(B^0 \to \phi K_1(1400)^0)}
&=&
\left\{ \begin{array}{ll}
 0.23^{+0.20}_{-0.16} \;\;\;\;\;\;\;\;{\rm \theta_{K_1} \sim 33^\circ}& \vspace{0.1cm} \\
0.50^{+0.43}_{-0.32} \;\;\;\;\;\;\;\;{\rm \theta_{K_1} \sim 58^\circ} &  \\ \end{array} \right. \;,
\eeq
 \beq
 \frac{\tau_{B^0}}{\tau_{B^+}} \cdot
 \frac{Br(B^+ \to \phi K_1(1270)^+)}{BR(B^0 \to \phi K_1(1270)^0)}
 &=&
\left\{ \begin{array}{ll}
0.98^{+0.99}_{-0.71} \;\;\;\;\;\;\;\;{\rm \theta_{K_1} \sim 33^\circ}& \vspace{0.1cm} \\
0.93^{+0.76}_{-0.58} \;\;\;\;\;\;\;\;{\rm \theta_{K_1} \sim 58^\circ} &  \\ \end{array} \right. \;,
\eeq
\beq
\frac{\tau_{B^0}}{\tau_{B^+}} \cdot
 \frac{Br(B^+ \to \phi K_1(1400)^+)}{BR(B^0 \to \phi K_1(1400)^0)}
 &=&
\left\{ \begin{array}{ll}
1.03^{+0.84}_{-0.67} \;\;\;\;\;\;\;\;{\rm \theta_{K_1} \sim 33^\circ}& \vspace{0.1cm} \\
1.07^{+0.92}_{-0.71} \;\;\;\;\;\;\;\;{\rm \theta_{K_1} \sim 58^\circ} &  \\ \end{array} \right. \;,
\eeq
which could be used to further determine the mixing angle $\theta_{K_1}$
and will be tested by the future precision B meson experiments.

\item
As seen in Table~\ref{tab:DA-pm}, the annihilation
diagrams have been straightforwardly and explicitly evaluated in the
pQCD approach. Furthermore, one can easily find out the large annihilation
contributions in the considered $B \to \phi K_1$ decays.
Therefore, whether the annihilation effects to these
decay modes is important or not can be determined by the future
precise measurements experimentally, which will provide useful hints to
understand the annihilation decay mechanism in $B$ meson physics and
identify the reliability of investigations in these kinds of decays by
employing the pQCD approach.
Frankly speaking, these branching ratios for the $B \to \phi K_1$ decays
predicted in the pQCD approach suffer relatively large uncertainties from
the currently less constrained hadronic parameters of the strange axial-vector
$K_{1A}$ and $K_{1B}$ states, which needs further improvements
from future experiments.

\end{itemize}

%%%%%%%%%%%%%%%%%%%%%%%%%%%%%%%%%%%%%%%%%%%%%%%%%%%%%%%%%%%%%%%%%%%%%%%%%
%%***********************************************************************
\begin{table}[hbt]
\caption{ The decay amplitudes(in unit of $10^{-3}\; \rm{GeV}^3$) of the
$B^+ \to \phi K_{1A}^+$ and $B^+ \to \phi K_{1B}^+$ modes
with three polarizations in the pQCD approach,
where only the central values are quoted
for clarification.}
\label{tab:DA-pm}
 \begin{center}\vspace{-0.3cm}{\small
\begin{tabular}[t]{c|c|c|c|c|c|c|c|c}
\hline  \hline
 Decay Amplitudes & ${\cal A}^T_{fe}$  & ${\cal A}^P_{fe}$
                  & ${\cal A}^T_{nfe}$ & ${\cal A}^P_{nfe}$
                  & ${\cal A}^T_{nfa}$ & ${\cal A}^P_{nfa}$
                  & ${\cal A}^T_{fa}$  & ${\cal A}^P_{fa}$\\
\hline \hline
  Channel   &  \multicolumn{8}{c}{$B^+ \to \phi K_{1A}^+$}\\
%   \hline
   \hline
 $L$      &$ 0.0$ &  $-4.51$
          &$ 0.0$ &  $\hspace{0.3cm} 0.45 - {\it i} 0.11$
          &$\hspace{0.3cm} 0.04 -{\it i} 0.02$
          &$\hspace{0.3cm} 0.01 - {\it i} 0.03$
          &$-0.05 +{\it i} 0.14$
          &$\hspace{0.3cm} 0.88 - {\it i} 0.91$
 \\
 $N$      &$ 0.0$ &  $-0.85$
          &$ 0.0$ &  $-0.32 - {\it i} 0.10$
          &$ \sim 0.00$ &  $-0.03 + {\it i} 0.03$
          &$\hspace{0.3cm} 0.15 + {\it i} 0.25$
          &$\hspace{0.3cm} 0.51 - {\it i} 2.13$
 \\
 $T$      &$ 0.0$ &  $-1.66$
          &$ 0.0$ &  $-0.71 - {\it i} 0.10$
          &$ -0.01 +{\it i} 0.01$   &  $ -0.06 +{\it i} 0.08$
          &$ -0.31 -{\it i} 0.51$ &  $ -1.92 - {\it i} 3.47$
 \\
 \hline \hline
  Channel  & \multicolumn{8}{c}{$B^+ \to \phi K_{1B}^+$} \\
\hline
 $L$      &$0.0$ &  $\hspace{0.38cm} 6.35$
          &$0.0$ &  $-0.23 + {\it i} 0.17$
          &$\hspace{0.38cm} 0.02 -{\it i} 0.06$ &  $-0.04 - {\it i} 0.04$
          &$-0.21 -{\it i} 0.23$ &  $-1.96 - {\it i} 0.67$
 \\
 $N$      &$ 0.0$ &  $\hspace{0.38cm} 0.64$
          &$ 0.0$ &  $\hspace{0.3cm} 0.22 + {\it i} 0.55$
          &$ \sim 0.00$ &  $ -0.03 + {\it i} 0.02$
          &$\hspace{0.3cm} 0.07 + {\it i} 0.05$  &  $ -0.19 - {\it i} 0.57$
 \\
 $T$      &$ 0.0$ &  $\hspace{0.38cm} 1.31$
          &$ 0.0$ &  $\hspace{0.3cm} 0.48 + {\it i} 1.11$
          &$ -0.01 +{\it i} 0.01$   &  $ -0.04 +{\it i} 0.07$
          &$ -0.14 -{\it i} 0.13$ &  $ -1.12 - {\it i} 0.61$
 \\
 \hline \hline
\end{tabular}}
\end{center}
\end{table}
%%%%%%%%%%%%%%%%%%%%%%%%%%%%%%%%%%%%%%%%%%%%%%%%%%%%%%%%%%%%%%%%%%%%%%%%%
%%***********************************************************************

\subsection{CP-averaged polarization fractions and relative phases}

Now we come to the analysis of the polarization fractions for  $B \to \phi K_1$ decays in the
pQCD approach. Based on the helicity amplitudes, we can define the
transversity amplitudes,
\beq
{\cal A}_{L}&=& \lambda m^{2}_{B}{\cal
M}_{L}, \quad
{\cal A}_{\parallel}=\lambda \sqrt{2}m^{2}_{B}{\cal
M}_{N}, \quad
{\cal A}_{\perp}=\lambda m_{\phi} m_{K_1}
\sqrt{2(r^{2}-1)} {\cal M }_{T}\;.
\label{eq:ase}
\eeq
for the longitudinal, parallel, and perpendicular polarizations,
respectively, with the normalization factor
$\lambda=\sqrt{G^2_{F}P_c/(16\pi m^2_{B}\Gamma)}$ and the ratio
$r=P_{2}\cdot P_{3}/(m_{\phi}\cdot
m_{K_1})$. These amplitudes satisfy the relation,
\begin{eqnarray}
|{\cal A}_{L}|^2+|{\cal A}_{\parallel}|^2+|{\cal A}_{\perp}|^2=1 %(\sum_i^{L,||,\perp} |{\cal A}_i|^2=1)\;.
\end{eqnarray}
following the summation in Eq.~(\ref{dr1}).
Since the transverse-helicity contributions manifest themselves in polarization
observables, we therefore define one kind of the polarization
observables, i.e., polarization fractions $f_{L,||,\perp}$
as,
\beq
f_{L,||,\perp}&=& \frac{|{\cal
A}_{L,||,\perp}|^2}{|{\cal A}_L|^2+|{\cal A}_{||}|^2+|{\cal
A}_{\perp}|^2},\label{eq:pf}
\eeq
With the above transversity amplitudes, the relative phases
$\phi_{\parallel}$ and $\phi_{\perp}$ can be defined as
 \beq
 \phi_{\parallel} &=& \arg\frac{{\cal A}_{\parallel}}{{\cal A}_L} \;,
   \qquad
 \phi_{\perp} = \arg\frac{{\cal A}_{\perp}}{{\cal A}_L} \;,
 \eeq

The theoretical results of polarization fractions and relative phases for these considered
$B \to \phi K_1$ decays in the pQCD approach have been displayed in Tables~\ref{tab:phik127p}-\ref{tab:phik1400}. Based on these numerical values,
some comments are given as follows:
\begin{itemize}
\item
Theoretically, the pQCD predictions of the longitudinal polarization fraction
$f_L$ for the $B^+ \to \phi K_1(1270)^+$ mode are
 \beq
 f_L(B^\pm \to \phi K_1(1270)^\pm)_{\rm pQCD} &=&
\left\{ \begin{array}{ll}
 0.47^{+0.31}_{-0.32}\;\;\;\;\;\;\;\;{\rm \theta_{K_1} \sim 33^\circ}& \vspace{0.1cm} \\
0.11^{+0.11}_{-0.07}\;\;\;\;\;\;\;\;{\rm \theta_{K_1} \sim 58^\circ} &  \\ \end{array} \right. \;,
 \eeq
Experimentally, the longitudinal polarization
fraction $f_L$ for the charged $B^+\to \phi
K_1(1270)^+$ decay is now available~\cite{Aubert:2008bc},
  \beq
 f_L(B^\pm \to \phi K_1(1270)^\pm)_{\rm Exp.} &=& 0.46^{+0.13}_{-0.15}\;,
  \eeq
It is obvious to see that the fraction with the smaller angle $\theta_{K_1} \sim 33^\circ$
is well consistent with the current data, which will be further examined by the
LHCb and/or Super-B measurements in the near future.

\item
In Refs.~\cite{Chen:2005cx,Cheng:2008gxa}, the authors have also evaluated
the polarization fraction of the $B^+ \to \phi K_1(1270)^+$ decays by employing
 GFA and QCDF, respectively.
However, it is noted that the longitudinal fraction
predicted in GFA is
$91.9\%(85.7\%)$~\cite{Chen:2005cx} with the mixing angle $\theta_{K_1} \sim 37^\circ(58^\circ)$,
which, in terms of the central value, is almost two times larger than the measured one.
As seen in Table~\ref{tab:phik127p}, the theoretical predictions for the longitudinal polarization fraction of $B^+ \to \phi K_1(1270)^+$ decay
in QCDF and pQCD approaches are consistent
with the current observation within still large errors.

\item
For other three $B \to \phi K_1$ decays, the longitudinal polarization fractions have also been
predicted in GFA, QCDF, and pQCD, respectively.
From the numerical results shown in Tables~\ref{tab:phik127p}-\ref{tab:phik1400},
it is interesting to find that the theoretical predictions of the longitudinal polarization fractions
for the $B \to \phi K_1(1270)$ decays are more sensitive than those for the $B \to \phi K_1(1400)$
decays to the variation of the mixing angle $\theta_{K_1}$ in both QCDF and pQCD approaches,
which is contrary to that observed in GFA~\cite{Chen:2005cx}:
$91.9\%(85.7\%)$ for $B \to \phi K_1(1270)$ decays and
$79.2\%(99.5\%)$ for $B \to \phi K_1(1400)$ decays with
the mixing angle $\theta_{K_1} \sim 37^\circ(58^\circ)$.
The above predictions and relevant phenomenologies will be tested by
future measurements at LHC and/or Super-B experiments.

\item
Up to now, there are no any available data and theoretical predictions on
the relative phases(in unit of rad) $\phi_{\parallel}$
and $\phi_{\perp}$ of the $B \to \phi K_1$ decays yet. It is therefore expected
that our predictions in the pQCD approach for the relative phases of these
considered $B \to \phi K_1$ decays as given in Tables~\ref{tab:phik127p}-\ref{tab:phik1400}
will be tested by the future LHCb and/or Super-B experiments.

\end{itemize}

\subsection{Direct CP-violating asymmentries}

Now we come to the evaluations of the CP-violating asymmetries of $B
\to \phi K_1$ decays in the pQCD approach.
For the charged $B$ meson decays, the direct CP violation
$\acp^{\rm dir}$ can be defined as,
 \beq
\acp^{\rm dir} =  \frac{|\overline{\cal A}_f|^2 - |{\cal A}_f|^2}{
 |\overline{\cal A}_f|^2+|{\cal A}_f|^2},
\label{eq:acp1}
\eeq
where ${\cal A}_f$ stands for the decay amplitude of $B^+ \to \phi K_1^+$,
while $\ov{{\cal A}}_f$ denotes the
charge conjugation one correspondingly. Using Eq.~(\ref{eq:acp1}),
we find the following pQCD predictions of the direct CP-violating asymmetries
\beq
\acp^{\rm dir}(B^+ \to \phi K_1(1270)^+)_{\rm pQCD}
&=&\left\{ \begin{array}{ll}
-0.7^{+3.6}_{-2.9}\times 10^{-2}\;\;\;\;\;\;\;\;{\rm \theta_{K_1} \sim 33^\circ}&  \vspace{0.1cm}\\
-1.3^{+1.1}_{-0.8}\times 10^{-2}\;\;\;\;\;\;\;\;{\rm \theta_{K_1} \sim 58^\circ} &  \\ \end{array} \right.
  \label{eq:adir-phik127p}\;,\\
\acp^{\rm dir}(B^+ \to \phi K_1(1400)^+)_{\rm pQCD}
&=&\left\{ \begin{array}{ll}
-1.5^{+1.2}_{-1.5}\times 10^{-2}\;\;\;\;\;\;\;\;{\rm \theta_{K_1} \sim 33^\circ}&  \vspace{0.1cm}\\
-1.3^{+1.5}_{-1.7}\times 10^{-2}\;\;\;\;\;\;\;\;{\rm \theta_{K_1} \sim 58^\circ} &  \\ \end{array} \right.
  \label{eq:adir-phik140p}\;;
\eeq
in which various errors
as specified previously have been added in quadrature. One can easily see that
the direct CP asymmetries of those two charged $B \to \phi K_1$ decays are around
$-3.6\% \sim +2.9\%$ ($-2.1\% \sim -0.2\%$) and $-3.0\% \sim -0.3\%$
($-3.0\% \sim +0.2\%$) with the mixing angle $\theta_{K_1} \sim 33^\circ (58^\circ)$,
respectively.
Note that these two channels exhibit much small direct CP-violating
asymmetries in the pQCD approach since the contributions coming
from the tree operators are approximately neglected in these two charged
$B^+ \to \phi K_1^+$ decays relative to the dominant penguin contributions,
which can be clearly seen from the decay amplitudes of every topology
as shown in Table~\ref{tab:DA-pm}.

At the experimental aspect, as mentioned in the Introduction,
the BABAR Collaboration has reported the measurements
of the direct CP violation for $B^\pm \to \phi K_1(1270)^\pm$ mode,
 \beq
A_{CP}^{\rm dir}(B^{\pm} \to \phi K_1(1270)^{\pm})_{\rm Exp.}&=&
(-15.0 \pm 20.0) \times 10^{-2}\;,
 \eeq
which is consistent with our pQCD calculations as shown in Eq.~(\ref{eq:adir-phik127p})
within errors.
It is worth of stressing that the preliminary measurements by BABAR
Collaboration suffer from large statistical and systematic errors.
One need more data from other experiments such as LHC and Super-B
to improve the precision of the direct CP asymmetry of $B^+ \to \phi K_1^+$ decays.

Meanwhile, by combining three
polarization fractions in the transversity basis with those of its
CP-conjugated $\bar B$ decays, we also computed the
direct CP violations of $B^+ \to \phi K_1^+$ decays in every polarization in the pQCD approach for
tests by future experimental measurements. The direct CP asymmetries of $B^+ \to
\phi K_1^+$ decays in the transversity basis can be defined as,
\beq
\acp^{\rm dir,\alpha}&=&
\frac{\bar f_\alpha- f_\alpha}{\bar f_\alpha+
f_\alpha}\;,
\eeq
where $\alpha=L,\parallel,\perp$  and  the definition of
$\bar f$ is same as that in~Eq.(\ref{eq:pf}) but for the
corresponding $\bar B$ decays. The numerical results for the
direct CP asymmetries of $B^+ \to \phi K_1^+$ decays in the transversity basis
within the framework of pQCD approach are presented in Table~\ref{tab:phik127p} and
\ref{tab:phik140p}, where the various errors as specified previously
have also been added in quadrature.

As for the CP-violating asymmetries for the neutral
$B^0 \to \phi K_1^0$ decays, the effects of
$B^0-\bar{B}^0$ mixing should be considered. However,
since they involve the pure penguin
contributions at leading order in the SM, which can be seen from the decay
amplitudes as given in Eq.~(\ref{eq:m0}), the considered two neutral modes
then present no direct CP violations in the SM.
If the measurements from experiments for the direct CP asymmetries
$\acp^{\rm dir}$ in $B^0 \to \phi K_1(1270)^0$ and $\phi K_1(1400)^0$ decays
exhibit large nonzero values,
which will indicate the existence of new physics beyond the SM and will provide a
very promising place to look for this exotic effect.

%%%%%%%%%%%%%%%%%%%%%%%%%%%%%%%%%%%%%%%%%%%%%%%%%%%%%%%%%%%%%%%%%%%%%%%%%
%%***********************************************************************
\begin{table}[hbt]
\caption{ The pQCD predictions for the CP-averaged branching ratios
and other physical observables for the considered $B \to \phi K_{1}$ decays with the inclusion
of the contributions from different sources as described in text.
Here only the central values are quoted for clarification with the mixing angle
$\theta_{K_1} \sim 33^\circ$.}
\label{tab:phys-t}
 \begin{center}\vspace{-0.3cm}{\small
\begin{tabular}[t]{c|c|c|c|c|c|c|c|c|c|c}
\hline \hline
     & decay rates
     & \multicolumn{3}{c|}{polarization fractions}
     &\multicolumn{2}{c|}{relative phases}
     &\multicolumn{4}{c}{direct CP asymmetries}\\
\hline  \hline
 Decay modes & BR$(10^{-6})$
             & $f_L$ & $f_{\parallel}$ & $f_\perp$
             & $\phi_{\parallel}$(rad) & $\phi_\perp$(rad)
             & $\acp^{\rm dir}$ & $\acp^{\rm dir,L}$
             & $\acp^{\rm dir,\parallel}$ & $\acp^{\rm dir,\perp}$\\
%   \hline
   \hline
 $B^+ \to \phi K_1(1270)^+$
         &$\begin{array}{cc} 4.7     \\ 7.5   \\ 5.3    \end{array}$
         &$\begin{array}{cc} 1.0     \\ 0.34 \\ 0.93  \end{array}$
         &$\begin{array}{cc} \sim 0.0\\ 0.39 \\ 0.04  \end{array}$
         &$\begin{array}{cc} \sim 0.0\\ 0.27 \\ 0.04  \end{array}$
         &$\begin{array}{cc}   \pi   \\ 2.19  \\ 4.49   \end{array}$
         &$\begin{array}{cc}   \pi   \\ 3.08  \\ 4.45   \end{array}$
         &$\begin{array}{cc}  0.0    \\ \sim 0.0  \\ 0.0   \end{array}$
         &$\begin{array}{cc}  0.0    \\ 0.05  \\ 0.0   \end{array}$
         &$\begin{array}{cc}  0.0    \\ -0.25   \\0.0     \end{array}$
         &$\begin{array}{cc}  0.0    \\ 0.29   \\0.0     \end{array}$
 \\   \hline
 $B^0 \to \phi K_1(1270)^0$
         &$\begin{array}{cc}   4.4     \\ 7.0  \\  5.0  \end{array}$
         &$\begin{array}{cc}   1.0     \\ 0.30 \\ 0.93  \end{array}$
         &$\begin{array}{cc}   \sim 0.0\\ 0.43 \\ 0.04  \end{array}$
         &$\begin{array}{cc}   \sim 0.0\\ 0.27 \\ 0.04  \end{array}$
         &$\begin{array}{cc}   \pi     \\ 2.33 \\ 4.50  \end{array}$
         &$\begin{array}{cc}   \pi     \\ 4.69 \\ 4.46  \end{array}$
         &$\begin{array}{cc}   0.0     \\  0.0 \\ 0.0   \end{array}$
         &$\begin{array}{cc}  0.0      \\  0.0 \\ 0.0   \end{array}$
         &$\begin{array}{cc}   0.0     \\  0.0  \\0.0     \end{array}$
         &$\begin{array}{cc}   0.0     \\  0.0  \\0.0     \end{array}$
 \\  \hline
 $B^+ \to \phi K_1(1400)^+$
         &$\begin{array}{cc}   32.0\\ 23.6 \\  30.4 \end{array}$
         &$\begin{array}{cc}   0.93\\  0.73\\  0.85  \end{array}$
         &$\begin{array}{cc}   0.04\\  0.12\\  0.08  \end{array}$
         &$\begin{array}{cc}   0.03\\  0.15\\  0.07  \end{array}$
         &$\begin{array}{cc}   \pi\\  4.24\\   3.37  \end{array}$
         &$\begin{array}{cc}   \pi\\  3.77\\   3.34  \end{array}$
         &$\begin{array}{cc}  0.0\\   -0.01\\   0.0 \end{array}$
         &$\begin{array}{cc}  0.0\\   -0.01\\   0.0 \end{array}$
         &$\begin{array}{cc}  0.0\\    -0.23\\  0.0   \end{array}$
         &$\begin{array}{cc}  0.0\\    0.14\\  0.0   \end{array}$
\\  \hline
 $B^0 \to \phi K_1(1400)^0$
         &$\begin{array}{cc}   29.6\\ 20.9 \\ 28.2  \end{array}$
         &$\begin{array}{cc}   0.93\\ 0.71 \\ 0.85  \end{array}$
         &$\begin{array}{cc}   0.04\\  0.13\\ 0.08  \end{array}$
         &$\begin{array}{cc}   0.03\\  0.16\\ 0.07  \end{array}$
         &$\begin{array}{cc}   \pi\\  4.31\\  3.38  \end{array}$
         &$\begin{array}{cc}   \pi\\  3.76\\  3.34  \end{array}$
         &$\begin{array}{cc}  0.0\\   0.0\\  0.0  \end{array}$
         &$\begin{array}{cc}  0.0\\   0.0\\  0.0  \end{array}$
         &$\begin{array}{cc}  0.0\\    0.0\\ 0.0    \end{array}$
         &$\begin{array}{cc}  0.0\\    0.0\\ 0.0    \end{array}$
 \\
 \hline \hline
\end{tabular}}
\end{center}
\end{table}
%%%%%%%%%%%%%%%%%%%%%%%%%%%%%%%%%%%%%%%%%%%%%%%%%%%%%%%%%%%%%%%%%%%%%%%%%
%%***********************************************************************

\subsection{Effects of annihilation contributions}
As discussed in
Ref.~\cite{Cheng:2008gxa}, the weak annihilation contributions play
a more important role in $B \to \phi K_1(1400)$ decays than that in
$B \to \phi K_1(1270)$ decays. 
At last, we will therefore explore the important contributions from the weak annihilation diagrams
to the penguin-dominated $B \to \phi K_1$ decays considered in this work.
In Tables~\ref{tab:phys-t} and \ref{tab:phys-t1}, we present the central values of
the pQCD predictions for the CP-averaged
branching ratios, the polarization fractions, the relative phases and the direct CP-violating
asymmetries with mixing angles $\theta_{K_1} \sim 33^\circ$ and $\theta_{K_1} \sim 58^\circ$
by taking the following three different sets of decay amplitudes into account:
\begin{enumerate}
\item[(1)] The factorizable emission diagrams only (the first entry);

\item[(2)] The factorizable emission  plus the weak annihilation contributions (the second entry);

\item[(3)] The factorizable emission plus the non-factorizable emission
contributions (the third entry).
\end{enumerate}

Then some phenomenological discussions are given as the following:

\begin{itemize}
\item
Generally speaking, by combining the analytic expressions as shown
in Eqs.~(\ref{eq:mp}-\ref{eq:m0}) and the numerical results of the decay amplitudes as presented in Table~\ref{tab:DA-pm} of $B^+ \to \phi K_1^+$ decays, it is clear to see 
that the $B^+ \to \phi K_1(1400)^+$
decay will be significantly dominated by the factorizable emission diagrams with both $\theta_{K_1}
\sim 33^\circ$ and $\theta_{K_1} \sim 58^\circ$, while the $B^+ \to \phi K_1(1270)^+$ decay
will be strongly determined by the annihilation diagrams with the increasing of the mixing angle
from $\theta_{K_1} \sim 33^\circ$ to $\theta_{K_1} \sim 58^\circ$.
These observations have been confirmed
through the central values of the CP-averaged branching ratios in the pQCD approach as
displayed in Tables~\ref{tab:phys-t}-\ref{tab:phys-t1}. Of course, the similar
phenomena will occur in the neutral $B^0 \to \phi K_1^0$ decays because of the negligible
contributions induced by the tree operators in the charged $B^+ \to \phi K_1^+$ decays.

\item
As far as the branching ratios are considered, one can see
from Table~\ref{tab:phys-t}-\ref{tab:phys-t1} that the annihilation diagrams
contribute to $B \to \phi K_1(1270)$ decays less(much larger)
than those to $B \to \phi K_1(1400)$ decays
with $\theta_{K_1} \sim 33^\circ(58^\circ)$.
More explicitly, without the annihilation contributions,
the branching ratios of $B \to \phi K_1(1400)$ modes become larger by about $21\% \sim 25\%$ with $\theta_{K_1} \sim 33^\circ$. However, by neglecting the weak annihilation contributions,
the branching ratios of $B \to \phi K_1(1270)$ decays decrease near $92\%$, while those
of $B \to \phi K_1(1400)$ decays increase around $63\% \sim 75\%$ with $\theta_{K_1} \sim 58^\circ$.

\item
For the polarization fractions and relative phases, one can also see that
the annihilation contributions play an important role in all the considered $B \to \phi K_1$ decays.
It is interesting to find that, analogous to $B \to \phi K^*$ decays, the weak
annihilation contributions could also reduce the longitudinal
polarization of the $B \to \phi K_1$ decays significantly.
While the non-factorizable emission diagrams play a minor role for these
quantities.

\item
Moreover, as claimed in the pQCD approach, the annihilation diagrams
provide the origin of the strong phases for predicting the CP violation
in the considered $B \to \phi K_1$ decays in the present work, which can be
seen obviously from the numerical values presented in Tables~\ref{tab:phys-t} and \ref{tab:phys-t1}.
Of course, the above general expectation for the pQCD approach
will be examined by the relevant experiments in the future, which could be
helpful to understand the annihilation decay mechanism in vector-vector
and vector-axial-vector $B$ decays in depth.

\end{itemize}

%%%%%%%%%%%%%%%%%%%%%%%%%%%%%%%%%%%%%%%%%%%%%%%%%%%%%%%%%%%%%%%%%%%%%%%%%
%%***********************************************************************
\begin{table}[hbt]
\caption{ Same as Table~\ref{tab:phys-t} but with mixing angle $\theta_{K_1} \sim 58^\circ$.}
\label{tab:phys-t1}
 \begin{center}\vspace{-0.3cm}{\small
\begin{tabular}[t]{c|c|c|c|c|c|c|c|c|c|c}
\hline \hline
     & decay rates
     & \multicolumn{3}{c|}{polarization fractions}
     &\multicolumn{2}{c|}{relative phases}
     &\multicolumn{4}{c}{direct CP asymmetries}\\
\hline  \hline
 Decay modes & BR$(10^{-6})$
             & $f_L$ & $f_{\parallel}$ & $f_\perp$
             & $\phi_{\parallel}$(rad) & $\phi_\perp$(rad)
             & $\acp^{\rm dir}$ & $\acp^{\rm dir,L}$
             & $\acp^{\rm dir,\parallel}$ & $\acp^{\rm dir,\perp}$\\
%   \hline
   \hline
 $B^+ \to \phi K_1(1270)^+$
         &$\begin{array}{cc} 0.4     \\ 10.4   \\ 0.7    \end{array}$
         &$\begin{array}{cc} 0.30     \\ 0.11 \\ 0.03  \end{array}$
         &$\begin{array}{cc} 0.41    \\ 0.47 \\ 0.52  \end{array}$
         &$\begin{array}{cc} 0.29    \\ 0.42 \\ 0.44  \end{array}$
         &$\begin{array}{cc}   \pi   \\ 3.59  \\ 2.76   \end{array}$
         &$\begin{array}{cc}   \pi   \\ 2.96  \\ 2.68   \end{array}$
         &$\begin{array}{cc}  0.0    \\ -0.01  \\ 0.0   \end{array}$
         &$\begin{array}{cc}  0.0    \\ 0.15  \\ 0.0   \end{array}$
         &$\begin{array}{cc}  0.0    \\ -0.26   \\0.0     \end{array}$
         &$\begin{array}{cc}  0.0    \\ 0.23   \\0.0     \end{array}$
 \\   \hline
 $B^0 \to \phi K_1(1270)^0$
         &$\begin{array}{cc}   0.4     \\ 10.4  \\  0.7  \end{array}$
         &$\begin{array}{cc}   0.29     \\ 0.12 \\ 0.03  \end{array}$
         &$\begin{array}{cc}   0.41     \\ 0.48 \\ 0.53  \end{array}$
         &$\begin{array}{cc}   0.29     \\ 0.40 \\ 0.44  \end{array}$
         &$\begin{array}{cc}   \pi     \\ 3.68 \\ 2.76  \end{array}$
         &$\begin{array}{cc}   \pi     \\ 2.95 \\ 2.67  \end{array}$
         &$\begin{array}{cc}   0.0     \\  0.0 \\ 0.0   \end{array}$
         &$\begin{array}{cc}  0.0      \\  0.0 \\ 0.0   \end{array}$
         &$\begin{array}{cc}   0.0     \\  0.0  \\0.0     \end{array}$
         &$\begin{array}{cc}   0.0     \\  0.0  \\0.0     \end{array}$
 \\  \hline
 $B^+ \to \phi K_1(1400)^+$
         &$\begin{array}{cc}   36.3   \\ 20.8 \\  34.9 \end{array}$
         &$\begin{array}{cc}   0.95   \\  0.90\\  0.88  \end{array}$
         &$\begin{array}{cc}   0.03   \\  0.04\\  0.07  \end{array}$
         &$\begin{array}{cc}   0.02  \\  0.06\\  0.06  \end{array}$
         &$\begin{array}{cc}   \pi   \\  3.99\\   3.48  \end{array}$
         &$\begin{array}{cc}   \pi  \\  3.75\\   3.46  \end{array}$
         &$\begin{array}{cc}  0.0   \\   -0.01\\   0.0 \end{array}$
         &$\begin{array}{cc}  0.0    \\   -0.01\\   0.0 \end{array}$
         &$\begin{array}{cc}  0.0   \\    -0.13\\  0.0   \end{array}$
         &$\begin{array}{cc}  0.0   \\    0.06\\  0.0   \end{array}$
\\  \hline
 $B^0 \to \phi K_1(1400)^0$
         &$\begin{array}{cc}   33.6\\ 17.7 \\ 32.4  \end{array}$
         &$\begin{array}{cc}   0.95\\ 0.89 \\ 0.88  \end{array}$
         &$\begin{array}{cc}   0.03\\  0.04\\ 0.07  \end{array}$
         &$\begin{array}{cc}   0.02\\  0.07\\ 0.05  \end{array}$
         &$\begin{array}{cc}   \pi\\  4.06\\  3.49  \end{array}$
         &$\begin{array}{cc}   \pi\\  3.76\\  3.46  \end{array}$
         &$\begin{array}{cc}  0.0\\   0.0\\  0.0  \end{array}$
         &$\begin{array}{cc}  0.0\\   0.0\\  0.0  \end{array}$
         &$\begin{array}{cc}  0.0\\    0.0\\ 0.0    \end{array}$
         &$\begin{array}{cc}  0.0\\    0.0\\ 0.0    \end{array}$
 \\
 \hline \hline
\end{tabular}}
\end{center}
\end{table}
%%%%%%%%%%%%%%%%%%%%%%%%%%%%%%%%%%%%%%%%%%%%%%%%%%%%%%%%%%%%%%%%%%%%%%%%%
%%***********************************************************************

%%%--=================================================================
%%%=====            Conclusions&Summary     ==========================
%%5===================================================================
\section{Conclusions and Summary} \label{sec:summary}

In this work, we studied the charmless hadronic $B \to \phi K_1$ decays,
which are dominated by the penguin contributions, by employing the pQCD
approach based on the framework of $k_T$ factorization theorem.
By taking the mixing angles $\theta_{K_1} \sim 33^\circ$ and $\theta_{K_1} \sim 58^\circ$  between the
two axial-vector $K_1(1270)$ and $K_1(1400)$ mesons, we explored the physical observables such as the CP-averaged branching ratios, the polarization fractions, the
relative phases, and the CP-violating asymmetries of the considered decay modes.

From our numerical and phenomenological studies we found the following points:
\begin{enumerate}
\item[(a)]
The pQCD predictions for the branching ratio, the polarization fractions, and the direct
CP asymmetry of the $B^\pm \to \phi K_1(1270)^\pm$ decay with the mixing angle
$\theta_{K_1} \sim 33^\circ$ are in good agreement with the current data as reported
by the BABAR Collaboration, which suggests that the small mixing angle $\theta_{K_1} \sim 33^\circ$
is possibly more favored.

\item[(b)]
For $B^\pm \to \phi K_1(1400)^\pm$ decay, however, the pQCD  predictions for its decay rate
with two different mixing angles basically agrees with the ones in the QCDF approach
within still large theoretical errors, but much larger than the preliminary
upper limit set by the BABAR Collaboration, which will be tested by the LHCb
and forthcoming Super-B experiments. Of course, the numerical results in pQCD
approach are consistent with the available upper limit roughly in 2$\sigma$ errors.

\item[(c)]
At the theoretical aspect, only parts of the numerical results predicted in
every different method or approach can be accommodated by the preliminary data.
Once the measurements reported by BABAR Collaboration would be confirmed
by the future measurements, it will be of great interest and probably a challenge to
further understand the $K_1$ hadrons' QCD behavior and the mixing angle $\theta_{K_1}$ between
two axial-vector $K_{1A}$ and $K_{1B}$ states.

\item[(d)]
The theoretical estimations on the relative phases and direct CP-violating asymmetries of
penguin-dominant $B \to \phi K_1$ decays are given for the first time in the pQCD approach,
which can also be tested by the experimental measurements in the near future.

\item[(e)]
The weak annihilation contributions play an important role
in $B \to \phi K_1(1270)$ and $\phi K_1(1400)$ decays.

\end{enumerate}

The pQCD studies for the four $B \to \phi K_1$ decays will be helpful for us
to understand the mixing angle $\theta_{K_1}$, the underlying helicity structure of
the decay mechanism, even the possible new physics effects in this type of decays.
We believe that many pQCD predictions presented in this paper will be tested in the near future,
when precision experimental measurements  become available.

%%%--=================================================================
%%%=====            Acknowledgements        ==========================
%%5===================================================================

\begin{acknowledgments}
This work is supported by the National Natural Science
Foundation of China under Grants No.~11205072 and No.~11235005,
and by a project funded by the Priority Academic Program Development
of Jiangsu Higher Education Institutions (PAPD),
by the Research Fund of Jiangsu Normal University under Grant No.~11XLR38,
and by the Foundation of Yantai University
under Grant No. WL07052.
\end{acknowledgments}

%%====================================================================
%%%%%%%%%%%%%%%%%%%%    References  ==================================
%%%%==================================================================

\end{document}